\documentclass[12pt,preprint]{aastex}
\slugcomment{}
\shorttitle{}
\usepackage[figuresright]{rotating}
\shortauthors{Cieza et al.}
\begin{document}
\title{\emph{Spitzer} observations of the Hyades: \\
Circumstellar debris disks at 625\,Myr of age}

\author{Lucas Cieza\altaffilmark{1,2,3},
William D. Cochran\altaffilmark{1}, and
Jean-Charles Augereau\altaffilmark{4}}

\altaffiltext{1}{Astronomy Department and McDonald Observatory,
University of Texas, Austin, TX 78712}
\altaffiltext{2}{Now at the Institute for Astronomy, University of Hawaii at Manoa, Honolulu, HI 96822}
\altaffiltext{3}{\emph{Spitzer} Fellow}
\altaffiltext{4}{Laboratoire d'Astrophysique de Grenoble, CNRS, Universit\'e Joseph-Fourier, UMR 5571, Grenoble, France}

\begin{abstract}
We use the \emph{Spitzer} Space Telescope to search for infrared excess at
24, 70, and 160\,$\mu$m due to debris disks around a sample of 45 FGK-type
members of the Hyades cluster.
We supplement our observations with archival 24 and 70\,$\mu$m
\emph{Spitzer} data of an
additional 22 FGK-type and 11 A-type Hyades members in order to provide
robust statistics on the incidence of debris disks at 625\,Myr of age,
an era corresponding to the late heavy bombardment in the Solar System.
We find that none of the 67 FGK-type stars in our sample show evidence for
a debris disk, while 2 out of the 11 A-type stars do so.
This difference in debris disk detection rate is likely to be
due to a sensitivity bias in favor of early-type stars. The fractional disk
luminosity, $L_{DUST}/L_*$, of the disks around the two A-type
stars is $\sim$4$\times$10$^{-5}$, a level that is below the
sensitivity of our observations toward the FGK-type stars.
However, our sensitivity limits for FGK-type stars are able to
exclude, at the 2-$\sigma$ level, frequencies higher than 12\% and 5\%
of disks with  $L_{DUST}/L_* > $1$\times$10$^{-4}$  and
$L_{DUST}/L_* > $5$\times$10$^{-4}$, respectively.
We also use our sensitivity limits and debris disk models
to constrain the maximum mass of dust, as a function of distance 
from the stars, that could remain undetected around our targets. 

\end{abstract}


\section{Introduction}\label{intro}

Soon after IRAS discovered cold circumstellar disks around main-sequence
(MS) stars \citep{AuBeGi84}, it was realized that these disks could not be
made of primordial material. Followup CO observations \citep[e.g.][]{YaHaOm93}
showed that molecular gas was highly depleted around these disks.
Since, in the absence of gas, the survival time of dust 
due to dissipation processes such as radiation/wind pressure and the  
Poynting-Robertson effect is much shorter than the ages of MS stars,
these systems are believed to be debris disks where dust is continuously
replenished by collisions between planetesimals, the building blocks of
planets. Because of their probable connection with the formation of
planetary systems, debris disks  rapidly became  the subject of many studies.
However, IRAS was only sensitive enough to study bright nearby objects and
most of the pre-\emph{Spitzer}
statistics come from surveys performed by IRAS's successor, the ISO satellite.
\citet{HaDoJM01} studied 84 nearby ($\mathrm{d} < 25$\,pc) A,F,G, and K stars
for which ISO was sensitive to photospheric fluxes and detected 60\,$\mu$m
excess in $\sim$50\% of the stars younger than 400\,Myr and in 10\% of
the stars older than 400\,Myr. They suggest that this sudden decrease
in the fraction of stars with disks around 400\,Myr is related to the lifetime
of planetesimals that replenish the dust.
\citet{SpSaSi01} observed $\sim$150 pre-main sequence stars and young
main sequence stars and detected 60\,$\mu$m excess in $\sim$ 25\% of
the objects. Their observations do not confirm a sudden decrease in
the disk fraction around 400\,Myr but rather suggest a power law
relationship (index $\sim$ --2) between the age of the star and the
fractional dust luminosity, $L_{DUST}/L_*$.
They argue that such a power law naturally arises in collisionally
replenished debris disks.  The discrepancies in the results from
these ISO surveys can probably be traced back to the different
target selection criteria and observing strategies and to small
number statistics.  However, since conclusions from \citet{HaDoJM01}
and  \citet{SpSaSi01} were not totally consistent, ISO was unable
to provide a clear picture for the evolution of debris disks.

Fortunately, \emph{Spitzer}'s unprecedented sensitivity has recently
allowed many studies of large samples of pre-MS and MS stars in the mid- and
far-IR. These studies are rapidly providing important clues on the evolution
of debris disks. Rieke et al. (2005) studied over 250 A-type stars and 
concluded that, even though the incidence of 24 $\mu$m excess clearly decreases 
with increasing stellar age, a very large dispersion on the magnitude
of the IR-excesses is seen at every age. 
Su et al. (2006) finds that A-type stars show a similar qualitative 
behavior at 70 $\mu$m. However, at this wavelength, the IR excess declines 
more slowly with time than at 24 $\mu$m.  The decay timescale is 
estimated to be $\sim$150 Myr for the 24 $\mu$m  excess and $\sim$850 Myr for 
the 70 $\mu$m excess. Trilling et al. (2007) studied field 
FGK-type stars and found that $\sim$4$\%$ and $\sim$16$\%$ of
them  show detectable IR excesses at 24 and 70$\mu$m, respectively. 
These disk frequencies are roughly a 
factor of two lower than those found for A-type stars at the same 
wavelengths. They also find that solar-type stars have an almost 
constant disk frequency beyond 1 Gyr and hence conclude that the 
their debris disk decay timescale is significantly larger than 
for A-type star.  Gautier et al. (2007) studied a sample of nearby M-type stars 
but detected no significant far-IR excesses. However, given the sensitivity of 
their survey, they conclude that the average MIPS excesses, measured 
in photospheric flux units, of the M-type stars are at least a factor 
of four lower than those of solar-type stars 

The \emph{Spitzer} study of the Hyades presented in this paper is intended 
to provide  additional clues by giving robust statistics on the frequency of 
debris disks at 625\,Myr of age for a homogeneous sample of MS stars. 
At 46\,pc, the Hyades is the nearest star cluster to the Sun, and represent 
a sample of stars formed at the same epoch with the same heavy element abundance
($[\rm{Fe/H}] = 0.13 \pm 0.01$) \citep{PaSnCo03}.

The 625\,Myr age of the Hyades places them at an extremely
interesting era in the evolution of planetary systems. 
This age corresponds almost exactly to the era of the late heavy
bombardment (LHB) in our Solar System about 3.9\,Gyrs ago
\citep{TePaWa73}.
The cratering record of the Moon,
Mars, and Mercury all indicate that the inner planets experienced
intense bombardment by large bodies at that time. There is still intense
debate as to whether the LHB represented
merely the end of an exponential decrease in the impact rate from the
formation of the terrestrial planets \citep{We75,We77,NeIv94},
or was instead a short intense spike in the bombardment rate
when the Solar System was about 600\,Myr old \citep{Ry90,Cohen02}.
In either case, the LHB of our Solar System clearly
indicates that at an age of $\sim$600 My there was still a major debris disk
present that was undergoing a rapid evolution.  Large, asteroid-size bodies
had been built up during the early planet building era, but not all of
these bodies had been incorporated into the planets. 
At the time of the LHB,  these bodies were undergoing
an era of significant collisions with the inner planets,
and presumably with each other as well. 
These collisions would have generated large amounts of smaller particles,
ranging all of the way down to dust particles.  
If most planetary systems go through a LHB-type event at a similar age,  
it is quite reasonable to assume that other stellar systems might have
also similar remnant debris disks at $\sim$600 Myr of age. However,
it has been recently argued that events such that of the LHB can be triggered
by sudden dynamical interactions between planets after a long 
quiescent  period of time (Gomes  et al. 2005).  In that case, the incidence 
of debris disks on Hyades members is not expected to be significantly 
different from that of stars that are somewhat older (or younger) than them.

Here, we analyze deep MIPS  24 and 70 $\mu$m observations for a sample
of 78 Hyades stars, enough to provide robust statistics on the status of
debris disks at 625\,Myr of age. In Section ~\ref{oba_data},  we describe the
sample of Hyades stars, our observations, and the data reduction procedures.
In Section ~\ref{results}, we establish our disk identification criteria
and present our detection statistics. In Section ~\ref{Discussion}, we
compare our detection statistics to recent \emph{Spitzer} results,
use our sensitivity limits and debris disk models to constrain 
the maximum mass of dust that could remain undetected around our targets,
and discuss the implications of our results for debris disk evolution models 
and the late heavy bombardment in the Solar System.

\section{Observations}\label{oba_data}

\subsection{\emph{Spitzer} Sample and Observations}

The majority of the 78 targets discussed in this paper were observed 
at 3.6, 4.5, 5.8 and 8.0 $\mu$m with the Infrared Array Camera 
(IRAC) and at 24, 70, and 160\,$\mu$m with the Multiband Imaging Photometer for
\emph{Spitzer} (MIPS) as part of our General Observer (GO) program 3371.
This program contains 45 FGK-type Hyades members from the high-precision
radial velocity (RV) survey discussed by \citet{CoHaPa02} and \citet{PaCoHa04}.
We have also included MIPS 24 and 70\,$\mu$m  observations of Hyades
members from the FEPS \emph{Spitzer} Legacy Project (PID=148, 21 FGK-type
stars), and two Guaranteed Time Observation (GTO) programs (PID=40,
11 A-type stars and PID = 71, 1 K0 star). Even though the MIPS photometry
for some of the targets has already been discussed  in the context
of their respective programs \citep{SuRiSt06,MeHiBa06}, we have
retrieved the MIPS data from the \emph{Spitzer} archive and processed
them ourselves for consistency.  The Astronomical Observation Requests
(AORs) keys, Program IDs, spectral types from the literature,
and near-IR photometry (from 2MASS) for our entire sample
of Hyades stars are listed in Table~\ref{tab:sample}.

Given the extreme near-IR brightness of the sample, the IRAC observations 
of our GO program were acquired in the subarray mode (e.g., 0.02 sec 
$\times$ 64 frames). The IRAC data were requested to try to better 
constrain the stellar photospheres of our targets; however, they did not 
fulfill their intended purpose and were not utilized in our 
study (see Section 3.1).
All the MIPS observations discussed in this paper were obtained using
the MIPS pointed imaging mode. At 24 $\mu$m, 1 or 2 cycles of 3-sec
frames were enough to detect the stellar photospheres of the entire
sample with high signal to noise ratio (S/N $>$ 40).
At 70 $\mu$m, the observations were designed for the 1-$\sigma$
\emph{detector} sensitivities to match the predicted stellar
photospheres of the targets. Thus, in the absence of background
noise, an IR excess 4 times larger than the photospheric flux would be
detected at a  5-$\sigma$ level. However, at the distance of the
Hyades, the 70 $\mu$m  photospheric emission of most FGK-type stars
falls below the extragalactic confusion level (see Section 3.2)
and the observations become background limited beyond 24 $\mu$m for
most of the sample.

\subsection{Data Reduction}

We processed the IRAC data (available only for the GO program 3371) 
and MIPS 24\,$\mu$m data using the mosaicing and source extraction 
software c2dphot, which was developed as part of the  the \emph{Spitzer} Legacy
Project ``From Molecular Cores to Planet Forming Disks'' Evans et al. (2006).
This program is based on the mosaicking program MOPEX (MOsaicker
and Point source EXtractor), developed by the \emph{Spitzer} Science
Center (SSC) and on the source extractor program DoPHOT \citep{ScMaSa93}.
The  IRAC and MIPS  24\,$\mu$m measurements  for our entire sample are
listed in Tables ~\ref{tab:IRAC} and ~\ref{tab:MIPS}, respectively.

For the 70\,$\mu$m and 160\,$\mu$m data, we used MOPEX to create mosaiced
images.  We started from the SSC pipeline version S14.4 of the 
median-filtered BCDs (basic calibrated data), which are
optimized for point-source photometry. For each source, we created two
versions of the 70\,$\mu$m mosaic, one resampled to  8$\arcsec$\,pixels
(close to the original size of the pixels in the detector) and the
other resampled to 4$\arcsec$\,pixels. We used the former to obtain the
aperture photometry and the latter to  visually inspect the images
for background contamination (See section~\ref{res_70}). The 160\,$\mu$m
data were resampled to mosaics with 16$\arcsec$\,pixels.

For the 70\,$\mu$m data, we use an aperture of 16$\arcsec$ in radius and a sky
annulus with an inner and an outer radius of 48$\arcsec$ and 80$\arcsec$,
respectively.
From high S/N 70\,$\mu$m point source observations we derive a multiplicative
aperture correction, $AC$, of 1.8. 
This AC is in good agreement with the 1.74 value suggested by the
\emph{Spitzer} Science Center for observations with the same aperture size
and similar sky annulus\footnote{See htt/ss.spitzer.caltech.edu/mips/apercorr.}.
Thus, we calculate the observed flux,
F$_{70}$, as F$_{70}$=FA$_{70} \times AC$, where FA$_{70}$ is the flux within
the aperture. We estimate the 1-$\sigma$ photometric uncertainty as
$\sigma = AC \times$ RMS$_{SKY}$ $\times$ $n^{1/2}$, where $RMS_{SKY}$
is the flux RMS of the pixels in the sky annulus, and $n$ is the number
of pixels in our aperture. The 70\,$\mu$m measurements for our entire
sample are listed in Table~\ref{tab:MIPS}. For the 160\,$\mu$m data,
we used an aperture with a radius of  32$\arcsec$ and a sky annulus with
an inner and an outer radius of 48$\arcsec$ and 80$\arcsec$, respectively.
The fluxes and uncertainty were calculated in the same way as for the
70\,$\mu$m data, but adopting an aperture correction of 2.0, appropriate
for the size of the aperture and sky annulus used$^4$.
The 160\,$\mu$m measurements for the entire sample of FGK-type stars from
program ID=3371 are also listed in Table~\ref{tab:MIPS} (the 160\,$\mu$m
data are not available for the Hyades stars from the other programs).

\section{Results}\label{results}

\subsection{MIPS 24\,$\mu$m results}\label{res_24}

At 24\,$\mu$m, all of our targets are detected with very high
signal to noise ratios (S/N $\sim$50-300). In order to establish
whether or not our targets show IR-excess at a given wavelength, we
first need to estimate the expected photospheric fluxes at that wavelength.
We do so by normalizing NextGen Models \citep{HaAlBa99}, corresponding to
published spectral types of our Hyades stars, to the near-IR data from 2MASS
listed in Table~\ref{tab:sample}. We decided not to include the IRAC data 
in the normalization of the stellar photospheres for two reasons. First, 
the IRAC data are only available for the stars from program 3371, and second 
because we found that including IRAC fluxes does not provide a better 
photospheric constraint than using the 2MASS data alone.  This is probably because, even 
for bright sources with formal errors $<$1$\%$, there is a random  error floor to 
the best uncertainty possible with our IRAC and MIPS 24 $\mu$m observing techniques 
and data reduction process of $\sim$0.05 mag (Evans et al. 2006). The IRAC photometry for 
all the targets in our GO program 3371 is presented for completeness only (see Table 2).
The expected 24\,$\mu$m photospheric fluxes for our entire sample, obtained as 
described above, are listed in Table~\ref{tab:MIPS}.

\begin{figure}
\plotone{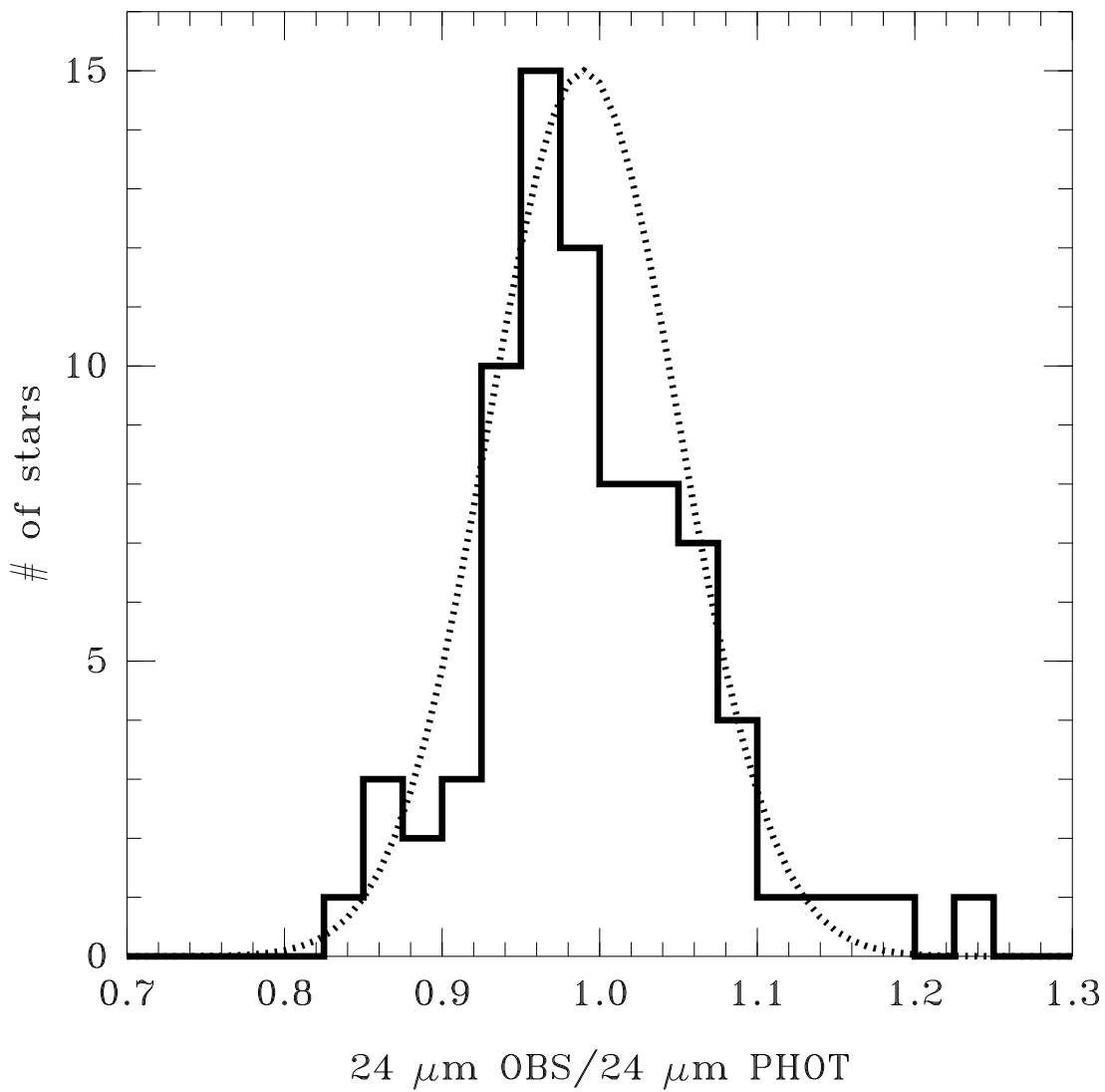}
\caption{Distribution of the observed 24\,$\mu$m fluxes
in units of the expected photospheric fluxes. A Gaussian
distribution centered at 0.99 and with a 1-$\sigma$
dispersion of 0.06 (dotted line)
is shown for comparison. Only one object, HD28355,
shows a significant ($>$ 3-$\sigma$) 24\,$\mu$m excess
above the predicted stellar photosphere.  HD28355
is an A-type star whose excess was already identified by
\citet{SuRiSt06}. \label{fig:24micron}}
\end{figure}

The magnitude of the smallest 24\,$\mu$m excess emission that we can
identify depends on both the uncertainty of our photometry and on
our ability to predict the photospheric flux. In Figure~\ref{fig:24micron},
we plot the distribution of observed 24\,$\mu$m fluxes relative to predicted
photospheric fluxes. After excluding a single outlier, this distribution
can be characterized as a Gaussian distribution with a mean of 0.99 and a
1-$\sigma$ dispersion of 0.06. The mean of the distribution is consistent with the 
absolute \emph{Spitzer} calibration uncertainty at 24 $\mu$m.
The dispersion of the distribution is identical to that found by \citet{BrBeTr06} 
for FGK-type field stars, 
smaller than that obtained by \citet{BeFrTr07} for FGK, and M stars, and only slightly 
larger than the uncertainty floor of 0.05 mag expected for 24 $\mu$m photometry obtained 
with \emph{c2dphot} (Evans et al. 2006). However, we note that our 1-$\sigma$ dispersion 
in the distribution of measured to predicted flux ratios is significantly larger 
than the $\sim$0.03 mag value achieved by Su et al. (2006) for A-type stars and 
by Thrilling et al. (2007) for FGK-type stars.
Based on the analysis of Figure~\ref{fig:24micron}, we conclude that only one of the Hyades stars
in our sample shows a significant ($>$ 3-$\sigma$) 24 $\mu$m excess.
This object is HD28355, an A-type star that was already identified by 
\citet{SuRiSt06} as having a debris disk. According to the values listed in
Table~\ref{tab:MIPS}, we find that the 24\,$\mu$m flux of HD28355 is
1.24 times the expected photospheric level, in good agreement with the
1.27 value found by \citet{SuRiSt06}.  

\subsection{MIPS 70\,$\mu$m results}\label{res_70}

\begin{figure}
\scalebox{.7}{\includegraphics{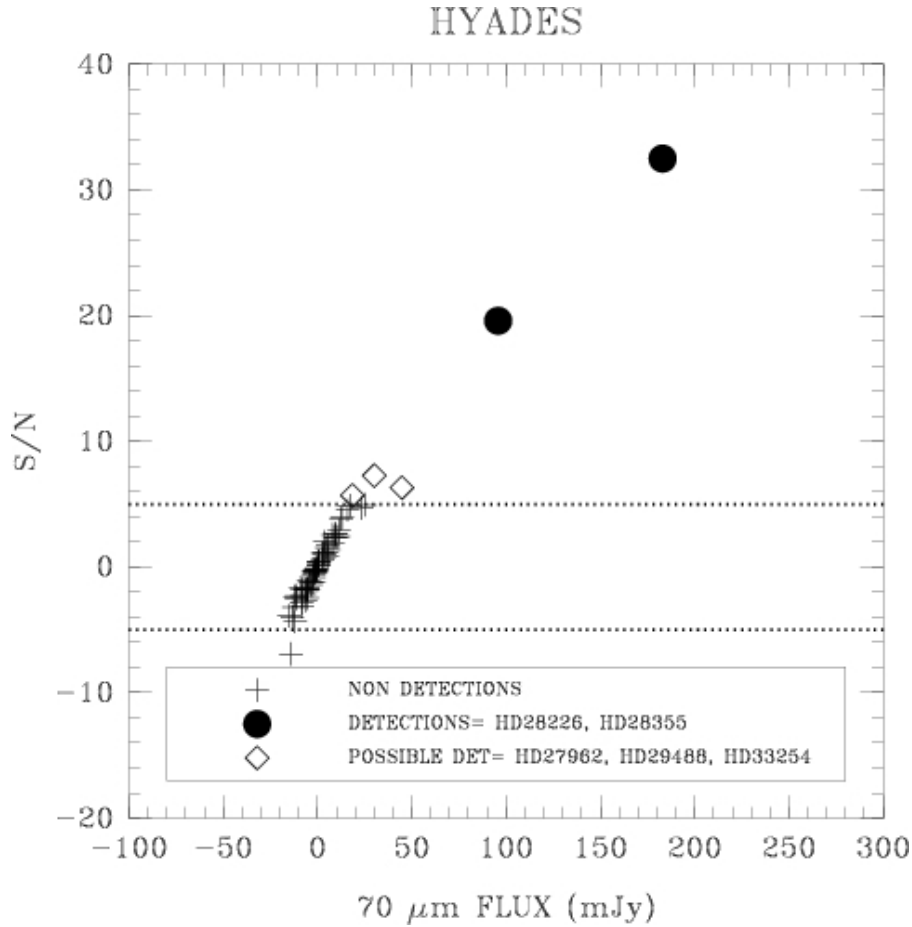}}
\caption{The signal to noise ratio versus the measured 70\,$\mu$m flux for
our sample of Hyades stars. Two objects shown as filled circles,
HD28226 and HD28355, clearly stand out as robust detections.
The horizontal dotted lines delimit the 5 $>$ S/N $>$ --5 interval.
For the deep 70\,$\mu$m observations considered in this paper, the noise
is dominated by the sky background variations. This variations are highly
none Gaussian and primarily due to extragalactic source confusion and
cirrus
contamination.  As a result, there is a similar number of objects
with S/N $\sim$ 5 and with S/N $\sim$ --5 and we consider objects S/N $<$ 5
to be none detections.
Three objects, HD27962, HD29488, and HD33524, have
S/N just above 5. We consider these objects to be possible
detections that need further consideration. \label{fig:SNR70micronflux}}
\end{figure}

\begin{figure}
\plotone{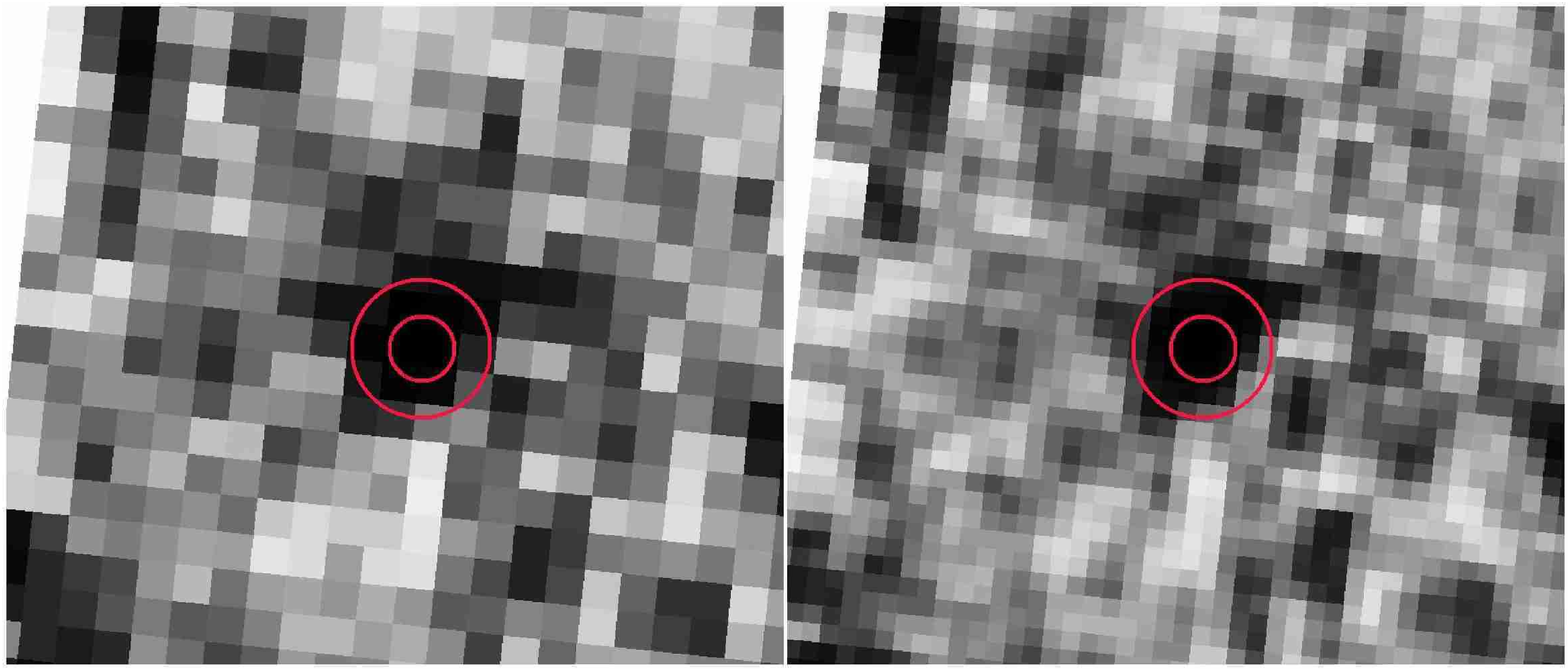}
\plotone{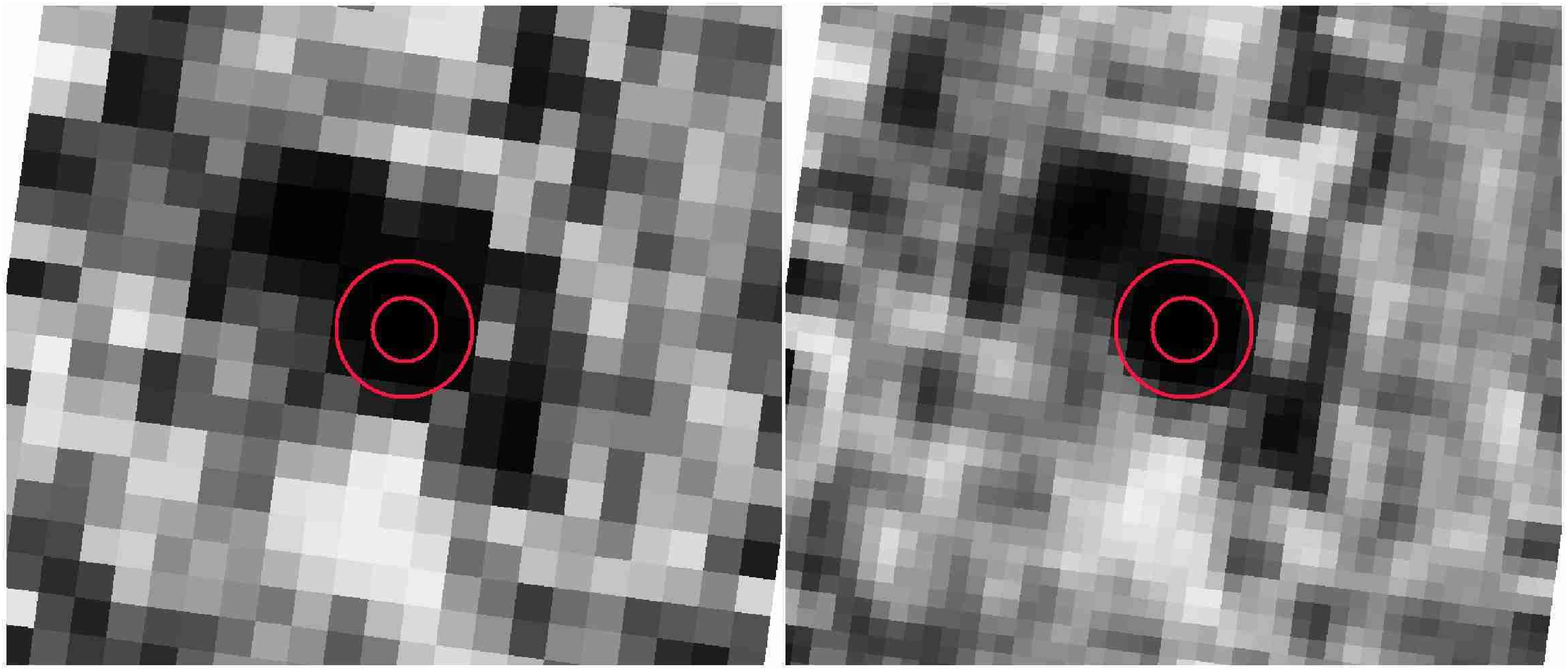}
\caption{The 70\,$\mu$m \emph{inverted} grayscale mosaics of HD28226 (top) and HD28355 (bottom).
North is up and East is to the left in all the images.
The images on the left correspond to the mosaics with 8$\arcsec$ pixels
(close to the physical size of the detector). The image on the right
correspond to the mosaics resampled to 4$\arcsec$ pixels
($\sim$half the physical size).
The 70\,$\mu$m emissions are centered at the positions of the objects
(marked by the concentric circles) and are detected with signal to noise 
ratios $\sim$20-30. Both objects are A-type stars already identified by
\citet{SuRiSt06} as having debris disks. \label{fig:HD28226}}
\end{figure}

By extrapolating the \emph{predicted} 24\,$\mu$m photospheric fluxes, listed 
in Table~\ref{tab:MIPS}, we estimated the fluxes of the 
stellar photosphereres that are expected at 70\,$\mu$m. These
predicted  70\,$\mu$m photospheric fluxes are also listed in 
Table~\ref{tab:MIPS}.
Unlike at 24\,$\mu$m, the expected photospheric flux at 70\,$\mu$m of all
our solar-type Hyades stars is at or below the noise of the observations. 
This noise is dominated by the sky pixel to pixel variations  due to
extragalactic source confusion and cirrus contamination. Bryden et al. 
(2006) present a detailed analysis of the sources of noise in deep 
70 $\mu$m observations. They analyze the noise of images constructed 
by combining an increasing number of cycles (each cycle consisting of 
ten 10-sec frames).  They show that beyond 4 cycles (a total exposure 
time of 400 sec), the total noise becomes dominated by the extragalactic
background noise. The low spatial resolution of \emph{Spitzer}, combined with the 
high instrumental sensitivity of MIPS, implies a high incidence of  
extragalactic sources per beam. 
This extragalactic source confusion sets a firm limit 
(1-$\sigma$ $\sim$2 mJy) to the sensitivity that can be achieved with MIPS at 70 
$\mu$m. This limit \emph{can not} be reduced with longer integration times. 
Any contamination from galactic cirrus decreases the sensitivity that can 
be achieved at any given field.  Most of the 70 $\mu$m observations discussed 
in this paper amount to at least 500 sec (and up to 1200 sec in many cases) 
and hence can be characterized as ``background limited''.
Since the background noise is highly non-Gaussian, a simple 3-$\sigma$
threshold is inappropriate to prevent spurious detections. 
Thus, the first step in our analysis  is to establish a different detection criterion.
In Figure~\ref{fig:SNR70micronflux}, we plot the signal to noise ratio
as a function of the measured 70\,$\mu$m flux. 
We find that a similar number of negative and
positive fluctuations exist at the 5-$\sigma$ level; therefore, we
consider objects with S/N ratios $<$\,5 to be non-detections.
We find that only two objects, HD28266 and HD28355, are unambiguously
detected.  In order to establish that the 70\,$\mu$m emission is in fact
associated with the Hyades targets, we inspect their mosaics, shown in
Figure~\ref{fig:HD28226}, and verify that the emission is centered
on the targets.  HD28266 and HD28355 are both A-type stars, which have
been already identified by \citet{SuRiSt06} as having a debris disk.
As mentioned in Section~\ref{res_24}, HD28355 also shows significant
24\,$\mu$m excess.

\begin{figure}
\plotone{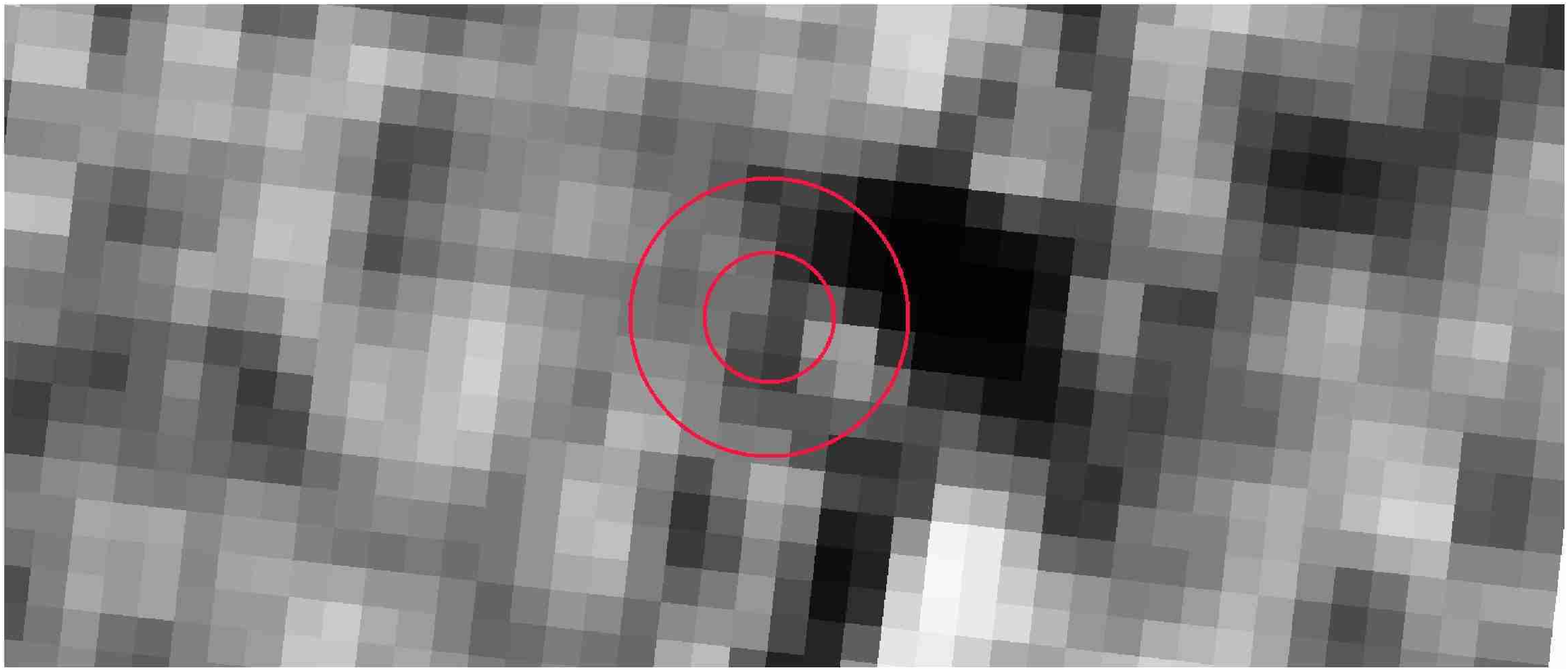}
\plotone{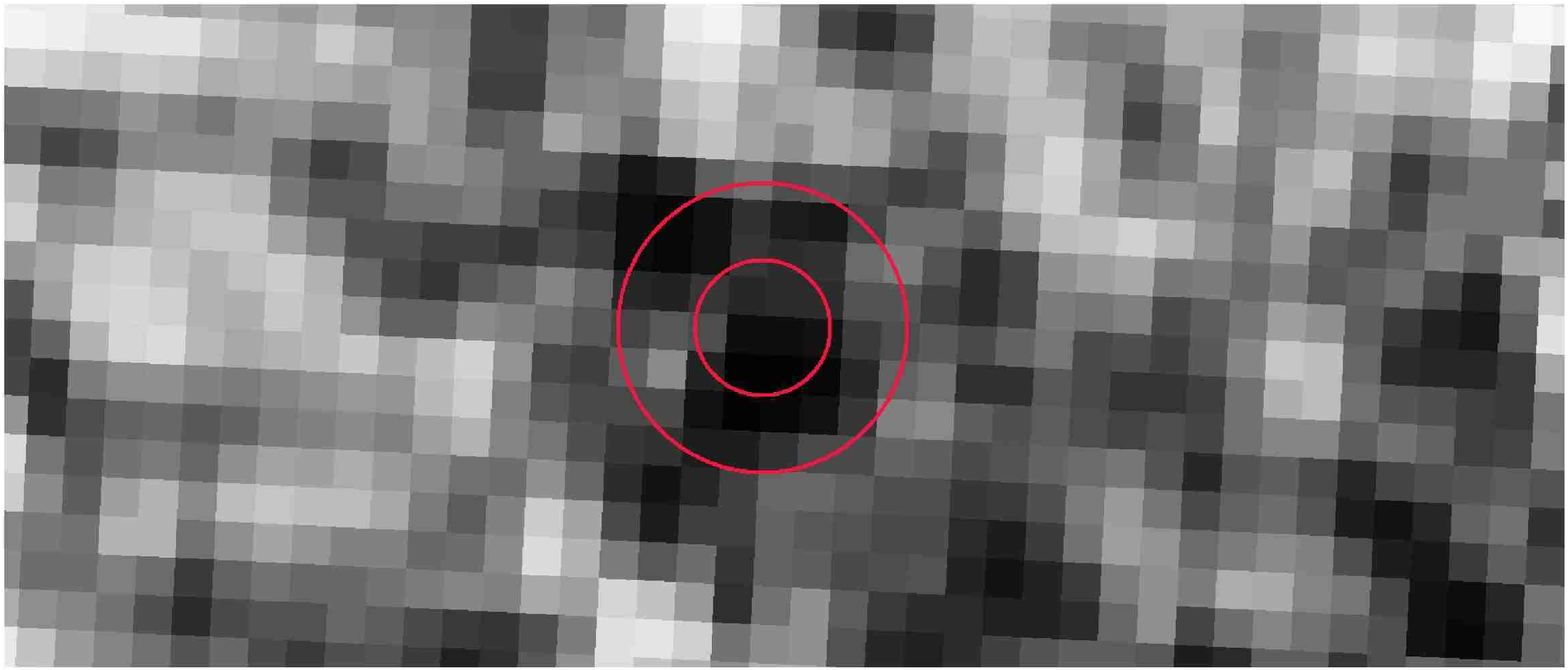}
\caption{The 70\,$\mu$m \emph{inverted} grayscale mosaics of HD27962 (top) and HD29488 (bottom)
resampled to 4$\arcsec$ pixels  in order to gain spatial resolution.
North is up and East is to the left in both images.
We find that 70 emission is detected at a S/N level $\sim$6-7 within the
our aperture (radius=16$\arcsec$). However, since the emissions are not
centered at the positions of the targets (marked by the
concentric circles), we conclude that they are not likely to be associated
with them. \label{fig:HD27962}}
\end{figure}

\begin{figure}
\plotone{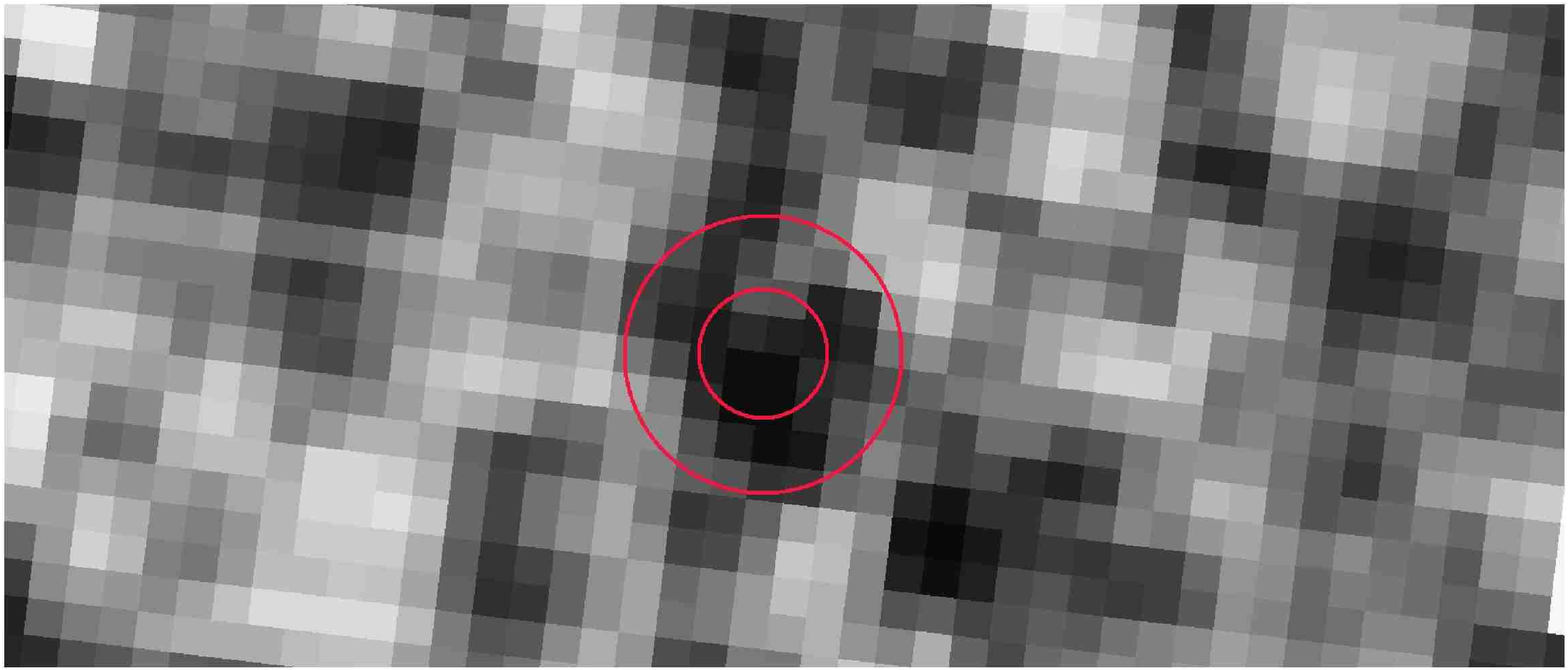}
\plotone{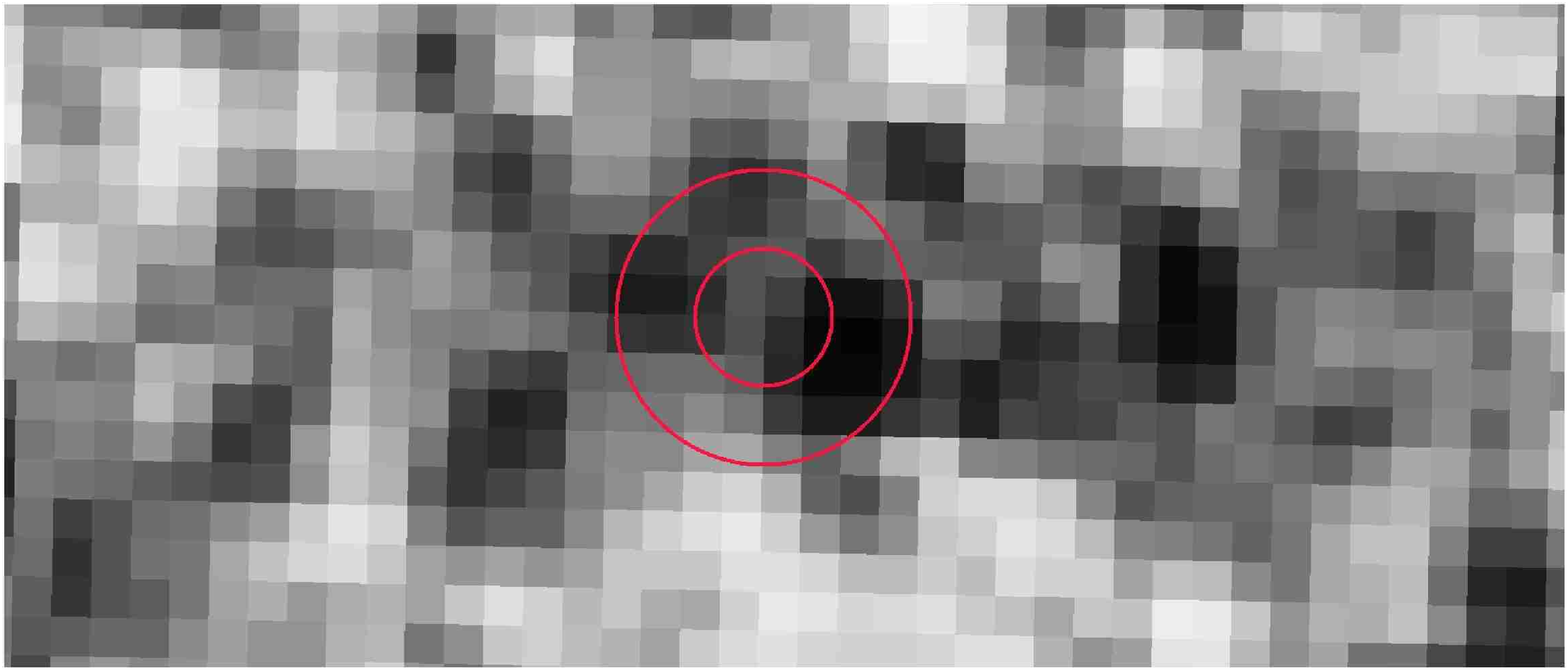}
\caption{The 70\,$\mu$m \emph{inverted} grayscale
mosaics of HD28527 (top) and HD33254 (bottom)
resampled to 4$\arcsec$ pixels  in order to gain spatial resolution.
North is up and East is to the left in both images.
Both objects are A-type stars  identified by \citet{SuRiSt06} as
having small 70\,$\mu$m excess (FLUX$_{70}$ /L$_{*,70}$ $\sim$2.7-2.9).
However, since we detect these objects at a marginal level
(S/N = 4.9 and 5.6, respectively) and the emissions are not
centered at the locations of the objects (marked by the concentric circles),
we do not consider the detections to be real. \label{fig:HD28527}}
\end{figure}

Three objects, HD27962, HD29488, and HD33524, have S/N just above 5.
We consider these objects to be possible detections that need further
consideration. We inspect their high resolution (4$\arcsec$ pixel) mosaics
(Figures~\ref{fig:HD27962} and \ref{fig:HD28527}) to establish the
spatial distribution of the 70\,$\mu$m emission.
In the three cases we find that even though there seems
to be a source near the aperture,  there is a significant offset
($\sim$10-15$\arcsec$) between the center of the 70\,$\mu$m emission and the
location of the Hyades target. Therefore, we conclude that the
70\,$\mu$m emission within these apertures are likely to be due to
background contamination.

\begin{figure}
\plotone{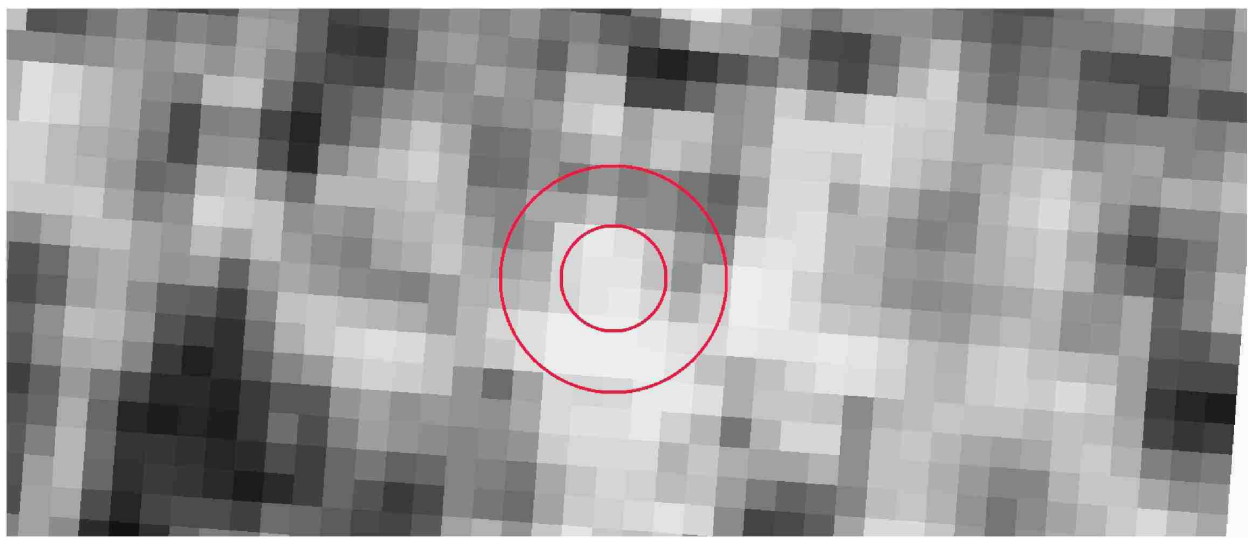}
\caption{The 70\,$\mu$m \emph{inverted} grayscale mosaic of HD28430 resampled 
to 4$\arcsec$ pixels. North is up and East is to the left.
Within the aperture centered at the target (marked by the concentric circles), there is a flux deficit
that is significant at the 6.9-$\sigma$ level. The existence of this
kind of minima strongly suggests that ``detections'' at the 5-7-$\sigma$
level should be interpreted with caution. \label{fig:HD28430}}
\end{figure}

We note that one of these objects,  HD33524, has been identified by
\citet{SuRiSt06} as having a weak 70\,$\mu$m excess. For this object, they
report a 70\,$\mu$m flux of $21.46\pm2.17$ mJy as opposed to our
$18.5\pm3.2$5 mJy (i.e., the fluxes agree very well within the uncertainties). 
A similar situation occurs for HD28527. \citet{SuRiSt06}
report a  70\,$\mu$m flux of  $37.36\pm5.94$  mJy, which also is in
relative agreement with  our $25.1\pm5.08$ measurement (within $\sim2\sigma$).
However, since the 70\,$\mu$m emission does not seem to be
centered at the target (Figure~\ref{fig:HD28527}), we do not consider
this detection to be real either. 
An independent reanalysis and inspection of the 70 $\mu$m data 
of HD 33254 and HD 28527 confirms that the 70 $\mu$m fluxes within
the apertures are no likely to be associated with the Hyades stars 
(Su 2008, private communications).
Our conservative detection criterion
is also supported by the presence of negative background fluctuations at 
the $5$-$7\sigma$ level at the location of some of the targets, such as HD28430 
shown in Figure~\ref{fig:HD28430}.

We conclude that none of the 67 FGK-type  Hyades stars in our sample are
detected at 70\,$\mu$m, while 2 of the 11 A-type stars are. The measured
70\,$\mu$m fluxes for these two objects, HD28266 and HD28355, are 13.9 and
11.5 times the values predicted for their respective photospheres
(see Table~\ref{tab:MIPS}).  We attribute these excesses, as \citet{SuRiSt06}
did, to the presence of debris disks around both of these sources.

\subsection{MIPS 160\,$\mu$m results}

\begin{figure}
\plotone{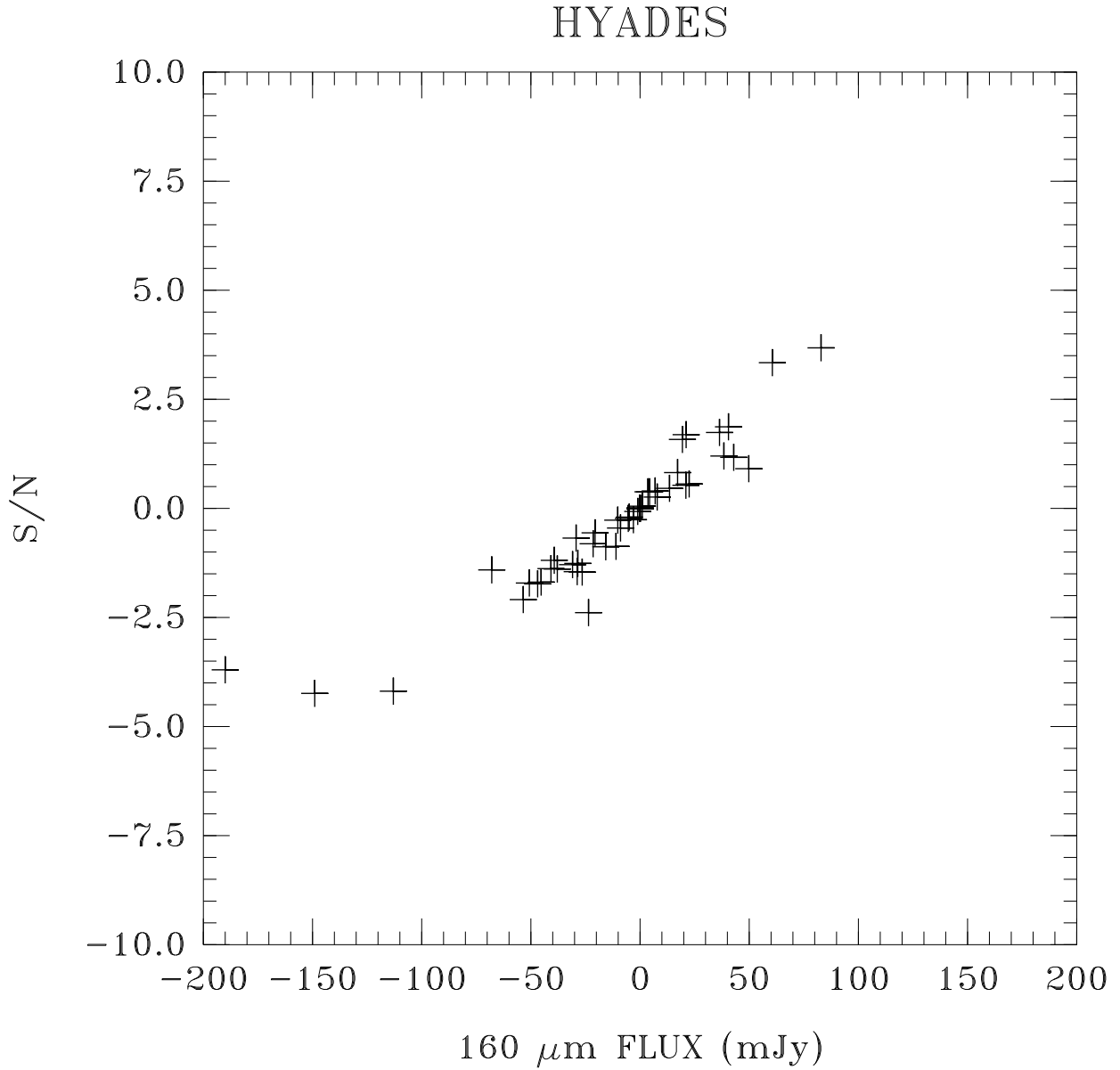}
\caption{
The signal to noise ratio versus the measured 160\,$\mu$m flux for
our sample of Hyades stars. We find no obvious 160\,$\mu$m detections.
At 160\,$\mu$m, as at 70\,$\mu$m, the noise is dominated by the extragalactic
source confusion and cirrus contamination. Thus, positive and negative
fluctuations at the $\sim$5-$\sigma$ level are not uncommon.
\label{fig:SNR160micronflux}}
\end{figure}

At 160\,$\mu$m, the expected photospheric levels are significantly below the
noise of the observations (which are only available for the 45 FGK-type
stars from the program ID=3371). 
In order to establish the detection of any of our targets, we follow the
same approach as for the 70\,$\mu$m data. In Figure~\ref{fig:SNR160micronflux},
we plot the  the signal to noise ratio as a function of the measured
160\,$\mu$m flux.  As for the 70\,$\mu$m observations, we find that a similar
number of negative and positive fluctuations exist at the 5-$\sigma$ level;
therefore, we considered all the 160\,$\mu$m measurements to be non-detections.

\section{Discussion}\label{Discussion}

\subsection{Comparison to Recent Spitzer Results}

Recent \emph{Spitzer} surveys have provided robust statistics on
the debris disk frequencies around nearby stars against which our
results can be compared. In order to make more meaningful
comparisons, we divide the sample into FGK-type stars (67 objects)
and A-type stars (11 objects). There are two motivations
for doing so. First, most of the previous studies are restricted
to either one of these groups. Second, given the strong luminosity
dependence on spectral type, the 70\,$\mu$m
observations are sensitive to much smaller 70\,$\mu$m excesses
(in units of photospheric fluxes) and  $L_{DUST}/L_*$ values
for A-type stars than for FGK-type stars.

\subsubsection{FGK-type vs A-type stars}\label{A-type}

\begin{figure}
\plotone{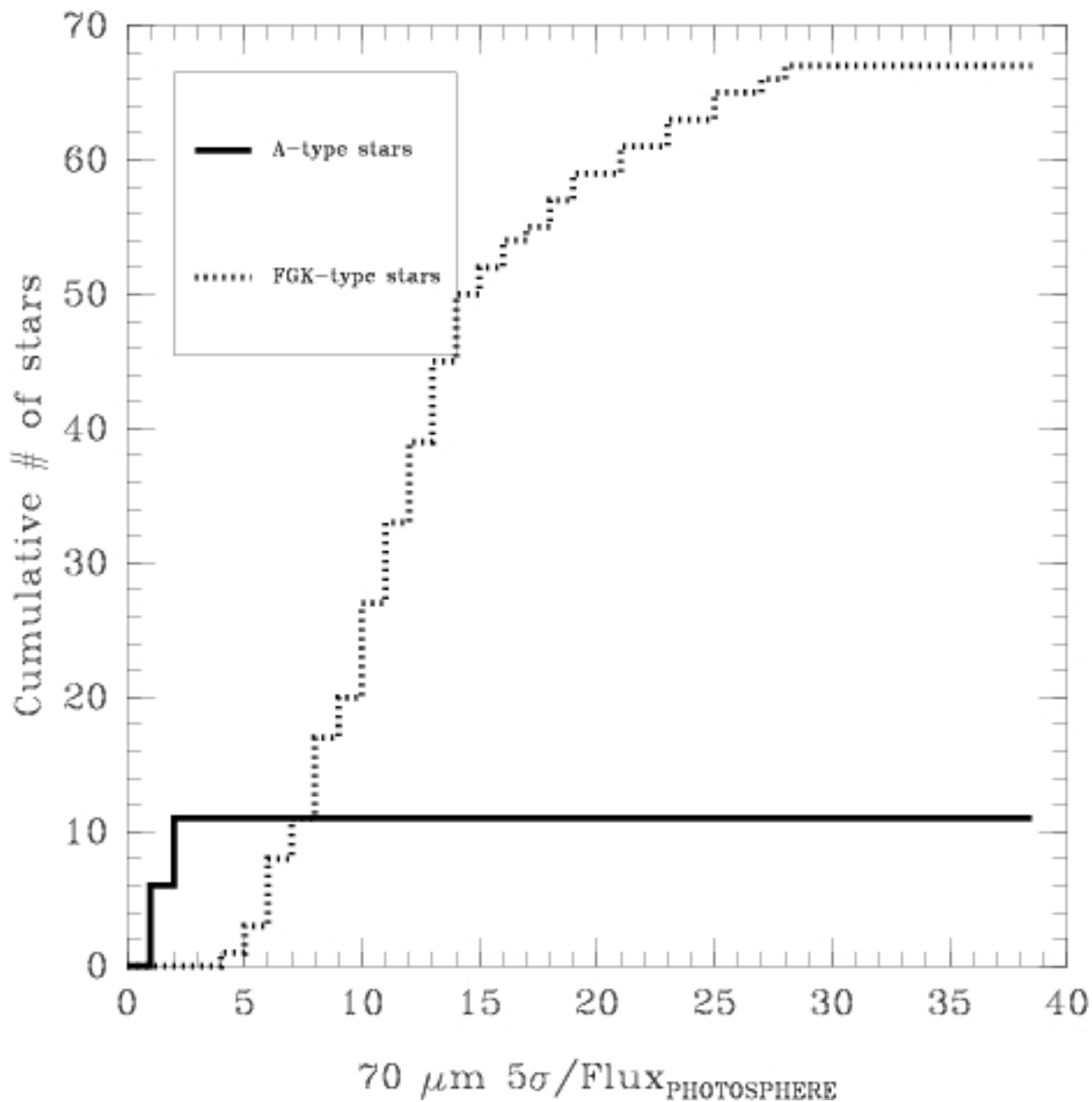}
\caption{Cumulative distribution of 70\,$\mu$m 5-$\sigma$
sensitivities in units of the expected photospheric fluxes for
FGK-type Hyades stars (dotted line) and A-type Hyades stars
(solid line). For A-type stars, the  70\,$\mu$m observations
can detect, at the 5-$\sigma$ level, fluxes that are
$\sim$1-2$\times$ that of the expected photospheres. In contrast,
for most of the FGK-type stars, the 70\,$\mu$m observations
are only sensitive enough to detect fluxes that are
$\sim$15$\times$ the expected photospheric values.
\label{fig:70micronCDF}}
\end{figure}

To estimate the sensitivity difference between the 70\,$\mu$m
observations of A-type stars and FGK-type stars, we calculate the
ratio of 5 times the flux uncertainties to the estimated photospheric
values (from Table~\ref{tab:MIPS}). A cumulative histogram of this
ratio is shown in Figure~\ref{fig:70micronCDF}
for FGK and A-type stars. For A-type stars, the 70\,$\mu$m observations
can detect fluxes that are $\sim$1-2$\times$ those of the expected
photospheres. However, for most of the FGK-type stars, the 70\,$\mu$m
observations are only sensitive enough to detect fluxes that are
$\sim$15$\times$ the expected photospheric values.
As discussed in Section 3.2, this sensitivity limitation 
is mostly due to the fact that the stellar photospheres of 
solar-type stars, at the distance of the Hyades, fall below the  
70 $\mu$m extragalactic confusion limit for MIPS 
(1-$\sigma$ $\sim$2 mJy, Bryden et al. 2006).

The difference in 70\,$\mu$m sensitivity is even larger when it is
calculated in terms of minimum detectable disk luminosity, $L_{DUST}/L_*$.
Following Bryden et al. (2006), we calculate minimum disk luminosity 
as a function of 70\,$\mu$m excess flux by setting the emission peak 
at 70\,$\mu$m (T$_{DUST} = 52.5 K)$, according to:

\begin{equation}\label{FDL}
\frac{L_{DUST}}{L_*}(minimum)
= 10^{-5}\left(\frac{5600 K}{T_{*}}\right)^3\frac{F_{DUST,70}}{F_{*,70}}
\end{equation}
where $F_{DUST,70}$ is the flux of the dust and $F_{*,70}$ is the flux of
the stars, both at 70\,$\mu$m.
By setting $F_{DUST,70} = 5\sigma_{70}-F_{*,70}$, we calculate the minimum
$L_{DUST}/L_*$ values that are detectable for A-type and FGK-type stars.
The results are shown in Figure~\ref{fig:70micronCDF2}, which demostrates that
the 70\,$\mu$m observations  of A-type stars are sensitive enough to detect
disk with $L_{DUST}/L_*$ values in the 10$^{-6}$-10$^{-5}$ range.
However for most of the FGK-type stars, the 70\,$\mu$m observations are only
sensitive enough to detect disks with 
$L_{DUST}/L_{STAR} \gtrsim$1-2$\times$10$^{-4}$.
We also use equation ~\ref{FDL} to estimate $L_{DUST}/{L_*}$ values
of 3.4$\times$10$^{-5}$ and 4.7$\times$10$^{-5}$ for the debris disks
around HD28226 and HD28355, respectively. 
Since there are only 2 FGK-type objects for which the 70\,$\mu$m
observations are sensitive enough to detect disks fainter than
$L_{DUST}/{L_*} \sim 4.0 \times$10$^{-5}$, we conclude that
the difference in the detection rate of debris disks around
A-type stars and FGK-type stars is the result of a
sensitivity bias rather than a real effect.

\subsubsection{Comparison to FGK-type field stars}

\citet{BrBeTr06} present 24 and 70\,$\mu$m observations for 69~FGK
nearby (distance $\sim$10-30\,pc) field stars with a median age of
$\sim$4\,Gyrs.
They find 24\,$\mu$m excess around only one of their targets. This is
consistent with the 24\,$\mu$m excess rate of 0$\%$ we find for our 67
FGK-type stars.
At 70\,$\mu$m, they identify 7 debris disks. This excess rate ($\sim$10$\%$),
if taken at face value, seems inconsistent with our results.
However, given the smaller distances to the stars involved, their 
survey was more sensitive than ours to faint disks.
Since the stellar photospheres of their targets are above 
the  70 $\mu$m confusion limit of MIPS, they were able to detect 
the stellar photosphere of most of them and identify very faint 
disks ($L_{DUST}/L_* \geq 10^{-5}$).
They also find that the disk frequency increases from 2$\%\pm2\%$
for disks with $L_{DUST}/L_*\geq 10^{-4}$ to $12\%\pm5\%$ for
disks with $L_{DUST}/L_* \geq 10^{-5}$. Recent results by Trilling et al. (2008) 
confirm that the frequency of debris disks with $L_{DUST}/L_* \geq 10^{-4}$ around 
solar type stars is $\sim$2$\%$.

\begin{figure}
\plotone{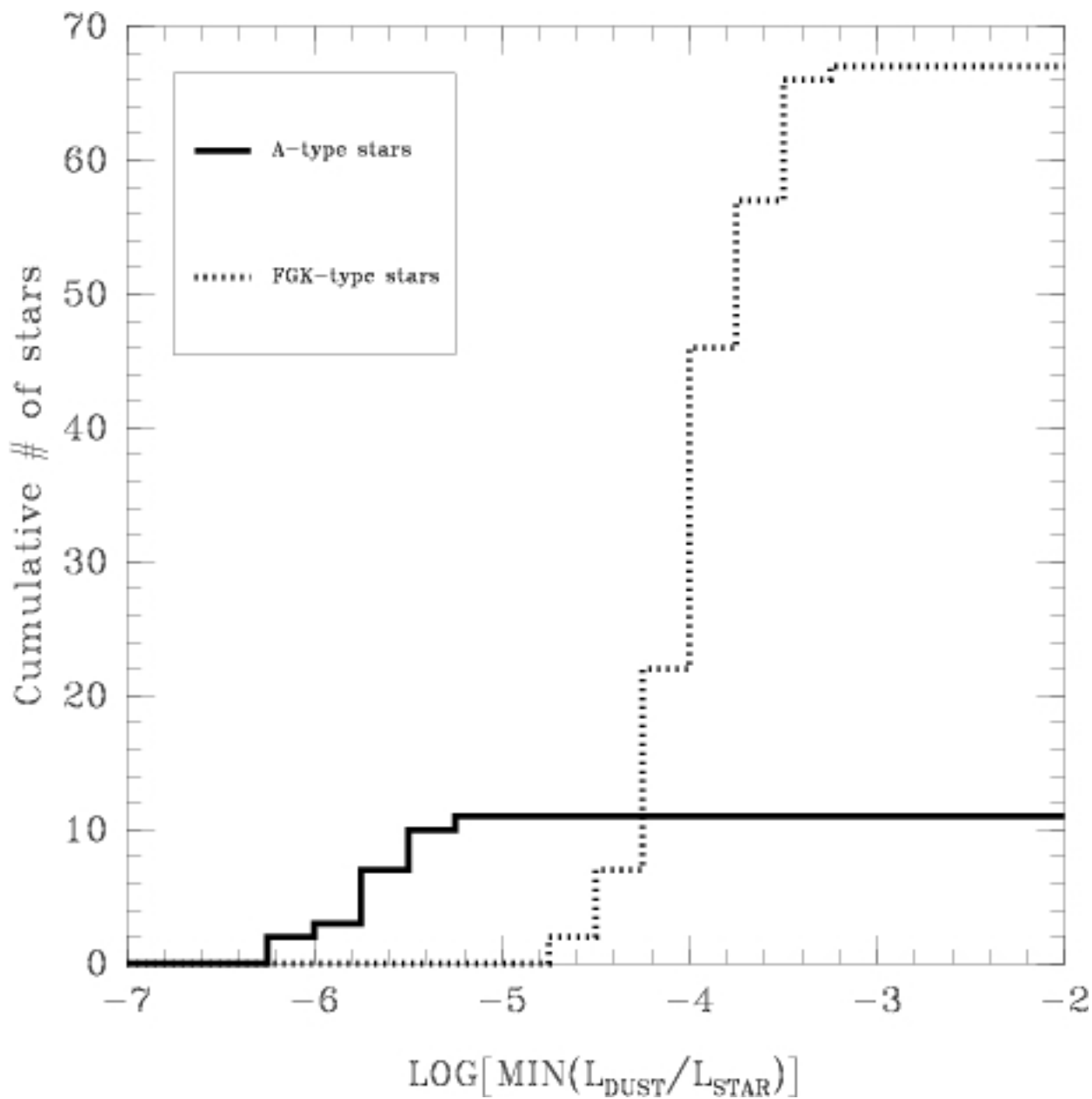}
\caption{Cumulative distribution of 70\,$\mu$m 5-$\sigma$
sensitivities translated into fractional disk luminosities
for FGK-type Hyades stars (dotted line) and A-type Hyades stars
(solid line). For A-type stars, the  70\,$\mu$m observations
are sensitive to disks with fractional disk luminosities,
L$_{DUST}$/L$_{STAR}$,
$\sim$5$\times$10$^{-6}$. In contrast,
for most of the FGK-type stars, the 70\,$\mu$m observations
are only sensitive enough to detect disks with
 L$_{DUST}$/L$_{STAR}$ $\gtrsim$ 2$\times$10$^{-4}$.
\label{fig:70micronCDF2}}
\end{figure}

Figure~\ref{fig:70micronCDF2} shows that there are only 22 objects for which
our 70\,$\mu$m observations are sensitive enough to
detect a debris disk with $L_{DUST}/L_* \geq 10^{-4}$.
We use binomial statistics to show that if the incidence of disks
brighter than $L_{DUST}/L_* = 10^{-4}$ is in fact 2\%
as found by \citet{BrBeTr06} and Trilling et al. (2007), 
there was a 26$\%$ probability that our survey would find zero disks. 
Also, given the cumulative
distribution of sensitivities shown in Figure~\ref{fig:70micronCDF2},
we use binomial statistics to calculate the minimum  disk frequencies,
as a function of $L_{DUST}/L_*$, that would be \emph{excluded}
at the 1- and 2-$\sigma$ level
(i.e., the disk frequencies that would give
our survey a 32\% and 5\% chance to result in zero detections).
These  disk frequencies are tabulated in Table~\ref{tab:disklimits}.

Based on the statistics for disks with $L_{DUST}/L_* \geq 10^{-4}$,
Table~\ref{tab:disklimits} suggests that a debris disk fraction in the Hyades
$\sim$2.5 and $\sim$6 times larger than in the field can be excluded at the
1-$\sigma$ and 2-$\sigma$ level, respectively.  Thus, we conclude that the
debris disk fraction of the FGK-type Hyades stars
(age $\sim$625\,Myr) is consistent with that in the field (age $\sim$~4~Gyrs),
but that $\sim$6$\times$ higher values cannot excluded from the currently
available data.

\subsection{Comparison to debris disk models}

In this section, we use the debris disk model developed by Augereau et al. (1999) 
to constrain the location and mass of the circumstellar material around
HD 28266 and HD 28355, the two A-type stars identified in section ~\ref{res_70}
as having real 70 $\mu$m excesses. We also use this model to estimate the 
maximum encompassed mass of dust, as a function of distance from the stars, that
could remain undetected around the A-type and FGK-type stars in our sample that
do not show significant IR excesses.

\subsubsection{The debris disks around HD 28266 and HD 28355}

We limit the exploration of the parameter space to the disk parameters that affect 
most the global shape of an SED, namely the minimum grain size $a_{\rm min}$, the peak surface 
density position $r_0$ and the total dust mass $M_{\rm dust}$ (or, equivalently, the 
surface density at $r_0$). 
We adopted a differential grain size distribution proportional to $a^{-3.5}$ between 
$a_{\rm min}$ and $a_{\rm max}$, with $a_{\rm max} = 1300\,\mu$m, a value sufficiently 
large to not affect the SED fitting in the wavelength range we consider.
Following Augereau et al. (1999), the disk surface density $\Sigma(r)$ is  parametrized by a 
two power-law radial profile 
$\Sigma(r) = \Sigma(r_0) \sqrt{2} \left(x^{-2\alpha_{\rm in}}+x^{-2\alpha_{\rm out}} \right)^{-1/2}$
with $x = r/r_0$, and where $\alpha_{\rm in} = 10$ and $\alpha_{\rm out} = -3$ to
simulate a disk peaked around $r_0$, with a sharp inner edge, and a density profile decreasing
smoothly with the distance from the star beyond $r_0$. The optical properties of the grains
were calculated for astronomical silicates (optical constants from Weingartner $\&$ Draine (2001)),
and with the Mie theory valid for hard spheres. The grain temperatures were obtained by assuming the
dust particles are in thermal equilibrium with the central star. NextGen model atmosphere spectra
(Hauschildt et al. 1999) scaled to the observed K-band magnitudes, were used to model the 
stellar photospheres.

\begin{figure}[tbp]
\centering
\includegraphics[angle=90,width=\columnwidth,origin=bl]{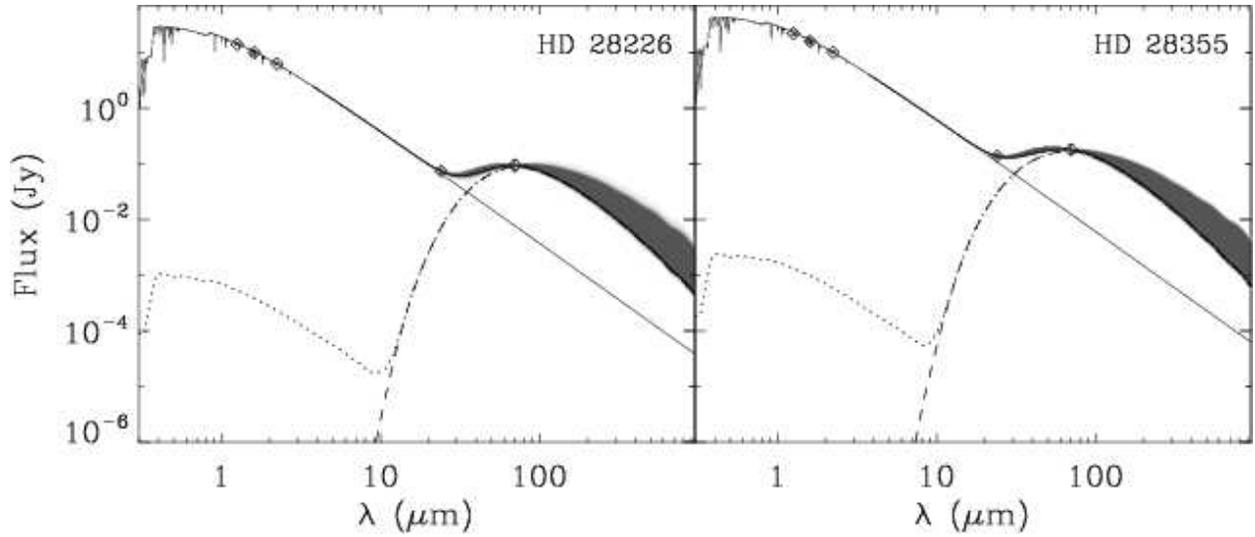}
\caption{Hyades A-type stars with 70 $\mu$m excess. On each plot, the darkest regions 
correspond to the most likely fits
to the SEDs. The dashed line shows the thermal emission for the best-fit model,
while the dotted line corresponds to the total disk emission (i.e. including scattered light emission).}
\label{SEDs}
\end{figure}

For each of the stars with 70 $\mu$m excess,  HD 28266 and HD 28355, we calculated $15000$ SEDs 
($0.3\,\mu$m$\leq \lambda \leq 950\,\mu$m), for $75$, logarithmically-spaced values of $a_{\rm min}$ 
between $0.05\,\mu$m and $100\,\mu$m, and for $200$ values of $r_0$, 
logarithmically-spaced between $10$\,AU and $500$\,AU. For each model, the dust mass 
was adjusted by a least-squares method, assuming purely photospheric emission in the 2MASS 
bands and by fitting the measured MIPS $24\,\mu$m and $70\,\mu$m flux densities.  
The results are summarized in Table 5, and the SEDs are displayed in Figure~\ref{SEDs}. 
Results in Table 7 are listed for two different regimes of minimal grain sizes, namely, 
$a_{\rm min}$ $>$ 10 $\mu$m and $a_{\rm min}$ $<$ 0.5 $\mu$m in order to illustrate the 
strong dependence of $r_0$ and M$_{dust}$ on the assumed grain size distribution. 

Even though neither the position of the peak surface density $r_0$, nor the minimum
grain size $a_{\rm min}$, can be uniquely determined with so few observational constraints,
some models can be eliminated. In particular, given the large luminosity of A-type 
stars, the small or nonexistent excess at 24 $\mu$m implies that the disks of 
HD 28226 and HD 28355 are significantly dust-depleted within \emph{at least} $\sim$40 AU from the star. 
However, ``inner-holes'' larger than 300 AU can not be excluded if
very small grains are present (i.e., a$_{\rm min}$ $<$ 0.5 $\mu$m). 
Similarly, even though the best-fit disk models imply  
dust masses of the order 10$^{-2}$M$_{\oplus}$, disk masses $\sim$0.1M$_{\oplus}$
could be accomodated for both stars. 

\subsubsection{Limits on dust masses as a function of radius}

In section ~\ref{A-type}, we found that the 70 $\mu$m observations of the Hyades 
were sensitive to 
significantly lower fractional disk luminosities for A-type stars than 
for FGK-type stars. However, since the 70 $\mu$m observations probe larger 
radii around A-type stars than around FGK-stars, this implies that the mass
of the grains needed to produce a given fractional luminosity, as calculated in 
section~\ref{A-type}, is \emph{larger} around A-type stars than it is around FGK-type 
stars (see Gautier et al. 2007 for a discussion of origin of the dependence of dust mass 
sensitivity on stellar luminosity).  Therefore, the degree to which the MIPS observations 
are sensitive to smaller amounts of dust around A-type stars than around FGK-type stars 
is not immediately obvious. In order to explore this last point, we estimate the maximum 
encompassed mass of dust, as a function of distance from the stars, that could remain 
undetected around the A-type and FGK-type stars in our sample that do not show 
significant IR excesses. 

\begin{figure}[h]
\plottwo{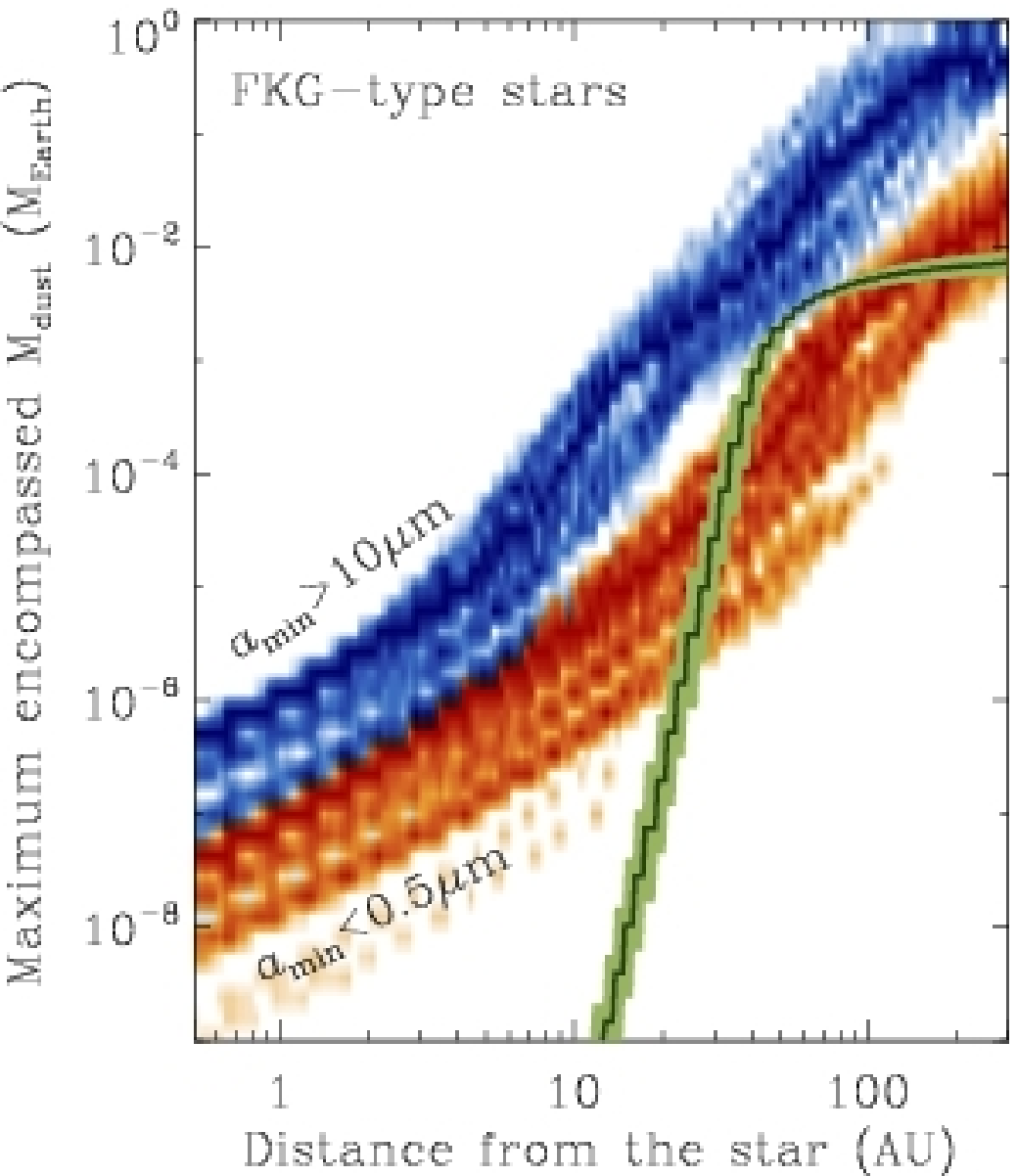}{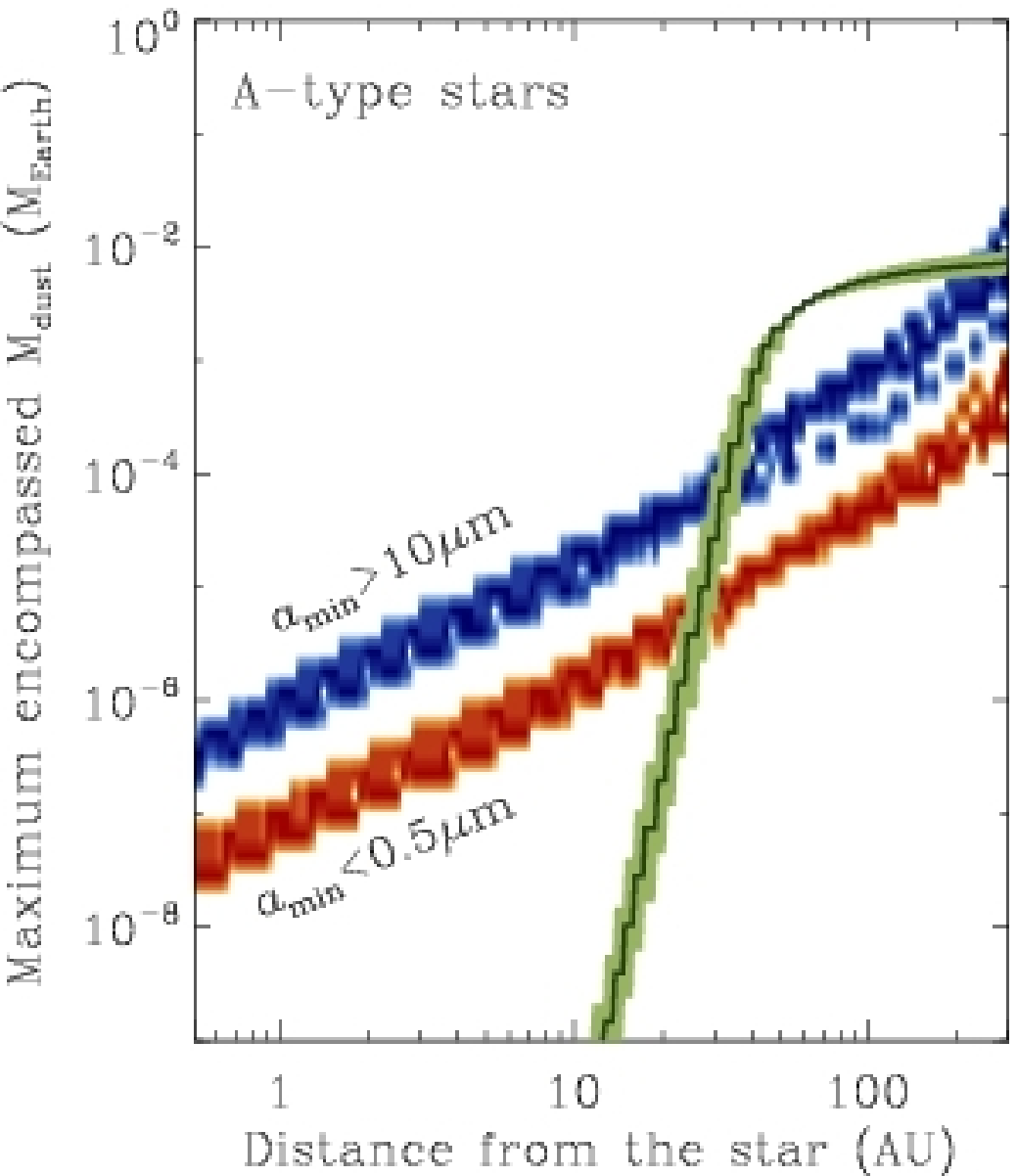}
\epsscale{1.5}
\caption{Maximum encompassed dust mass as a function of the distance from the star for
the FGK and A-type stars without \emph{Spitzer} excesses 
(respectively left and right panels). The red area corresponds to mass upper limits 
when minimum grain sizes $a_{\rm min}$ between $0.05\,\mu$m and $0.5\,\mu$m are 
considered, while the blue area corresponds to $10\,   \mu$m$  <  a_{\rm min}  
< 100\,   \mu$m. The solid black line corresponds to the mass as a function
of radius for the best-fit model of HD 28355 corresponding to the case where 
a$_{\rm min}$ $>$ 10 $\mu$m (see Table 5). The green region
indicates the 1-$\sigma$ limits of the best-fit model.}
\label{MvsR}
\end{figure}

Following Cieza et al. (2007), we use the optically thin disk models discussed above 
to constrain the maximum amount of dust that could be present within 300 AU of 
the A-type and FGK-type stars in our sample.  
Using the 70 $\mu$m 5-$\sigma$ upper limits (and the 160 $\mu$m limits, when available), 
we calculated $15000$ models analogous to those calculated for HD 28266 and HD 28355
for each of the stars in our sample. For each model, we calculated the mass encompassed 
within a radius $r$, as a function of this radius. With this approach, we 
estimate the maximum dust mass in the circumstellar regions of the Hyades stars with no 
detectable emission in excess to the photospheric emission. The results for A-type and 
FGK-type stars, for two different minimal grain size regimes (a$_{\rm min}$ $>$ 10 $\mu$m 
in blue and a$_{\rm min}$ $<$ 0.5 $\mu$m in red), are shown in Figure \,\ref{MvsR}.
This figure shows that the absence of IR excess at 24 and 70 $\mu$m constrains the total 
mass of dust within 10 AU of the solar type stars to be $<$ 10$^{-4}$ M$_{Earth}$.
This limit is an order of magnitude lower for A-type stars. 
The range of encompassed dust masses as a function of radius for the best-fit disk models 
of HD 28355 corresponding to the case where a$_{\rm min}$ $>$ 10 $\mu$m is shown for 
comparison (green region). The fact that, for FGK-type stars, the green region 
lies below the blue region (i.e., the case corresponding to the same grain size 
distribution), implies that disks similar to that found around HD 28355 would not 
be detectable by our observations if they were also present around the 
FGK-type Hyades members. Given the similarities of the inferred disk properties 
of HD 28355 and HD 28266 (see  Table 5), the same conclusion applies for HD 28266.
Thus, Figure \,\ref{MvsR} strengthen our conclusion from section ~\ref{A-type} stating that the 
difference in detection rate of debris disks around A-type and FGK-type stars 
is due to a sensitivity bias rather than to a real difference in the 
incidence or properties of debris disks around stars of different spectral types.

\subsection{Debris disk evolution and the Late Heavy Bombardment}

\subsubsection{Steady State vs. Stochastic Evolution}

\citet{RiSuSt05} studied a sample of 266 A-type stars with
\emph{Spitzer}, ISO, or IRAS 24 $\mu$m/25 $\mu$m data.
They used this very large sample to establish statistically significant
trends of IR-excess with age. They find that: (1) at all ages, the
population is dominated by stars with little or no IR excess,
(2) stars with a wide range of excesses are seen at every age,
and (3) both the frequency and the magnitude of the IR excess decreases
with time. In particular, they find that the upper envelope of the
evolution of the excess ratio with time can be fitted
by $t_o/t$, with $t_o\sim$150\,Myr. Similar trends are seen in the 70\,$\mu$m
excesses of A-type stars \citep{SuRiSt06},
with the difference that the decay time seems to be considerably larger,
$t_o \gtrsim$400\,Myr, suggesting inside-out disk clearing.

Based on these results, \citet{RiSuSt05}  argue that the evolution of
debris disks is the convolution of a stochastic and a steady component.
They suggest that, at any given age, the debris disks detected are those
that have experienced large planetesimal collisions in the
recent past. This stochastic evolution is on top of steady decrease in the
number of parent bodies in the belts  of planetesimals where the dust is
produced, which would explain the overall decrease of IR excess with age.
However, it has also been argued that a stochastic component in the evolution
of debris disk is not necessary to explain the diversity of disk properties
observed at a given age.
\citet{WySmSu07} construct a simple collisional model, where the mass
of planetesimals is constant until the largest ones reach collisional
equilibrium, at which point mass falls as 1/time.
They propose that the large spread in IR properties observed at
any given age can be explained in terms of the initial distributions
of masses and temperatures of the planetesimal belts
producing the dust. They argue that their simple model can account for the
24 and 70\,$\mu$m statistics presented by \citet{RiSuSt05} and
\citet{SuRiSt06} using realistic belt parameters, and thus that
transient events are not \emph{required} to explained the observations.
Given the limited observational constraints available, the models presented
by \citet{WySmSu07} do not rule out the possibility that
stochasticity plays an important role in the evolution
of most debris disks. Our results could provide additional constrains  to
these kinds of models because, unlike the studies by 
\citet{RiSuSt05} and \citet{SuRiSt06}, our study  provides
robust statistics for the debris disks at a single, well defined age.

\subsubsection{Implications for the Late Heavy Bombardment in the Solar
System}\label{LHB}

The 625\,Myr age of the Hyades corresponds almost exactly to the era of the
late heavy bombardment \citep[LHB,][]{TePaWa73,GoLeTs05}.
Thus, \emph{if} the Solar-type (FGK) Hyades stars resemble the Sun at
625\,Myr of age, our statistics could provide valuable clues
on this important event in the history of the Solar System.

The cause and the duration of the LHB is still a matter of debate.
Proposed causes for a intense spike in the impact rate include the
formation of Uranus and Neptune \citep{LeDoCh01}, the presence
of a fifth terrestrial planet in a low-eccentricity orbit which
became dynamically unstable at an age of about 600\,Myr
\citep{ChLi02}, or impacts by bodies left over from
planetary accretion \citep{MoPeGl01}.
More recently, \citet{GoLeTs05} propose that the LHB was triggered
by the sudden migration of the giant planets that occurred after
a long quiescent period of time.  In their model, soon after the
dissipation of the solar nebula, the orbits of Jupiter and Saturn
started to slowly diverge  due to the interaction with the massive
disk of planetesimals that was still present.  They argue
that $\sim$700\,Myr later, when Jupiter and Saturn crossed their 1:2 
mean motion resonance, their orbits became eccentric and temporally
destabilized those of Uranus and Neptune.  The reconfiguration of
the orbits of the giant planets resulted in the perturbation and
massive delivery of planetesimals to the inner Solar System, which
according to their models, lasted between 10-150\,Myr.

The observational signatures of a LHB-type event as seen from a distance of
46 pc are not known, but it as been suggested that they could be those of a
family of rare Solar-type stars  characterized by the presence of a bright
``hot disk'' around an object hundreds of million years old.
These objects present excess IR emission, originating in the
terrestrial planet regions, with $F_{DUST}/F_* > 10^{-4}$,
a level that is $>1000$ times larger than steady state evolution models
can explain \citep{WySmGr07}.
There are currently only 5 known objects that fall into this category of
``hot transient disks'':
BD~+20~307 \citep[age $\sim$300\,Myr,][]{SoZuWe05},
HD72905 \citep[age $\sim$400\,Myr,][]{BeBrSt06}),
$\eta$ Corvi \citep[age $\sim$1\,Gyr,][]{WyGrDe05},
HD69830 \citep[age $\sim$2\,Gyr,][]{BeBrRi05},
and $\tau$ Ceti (age $\sim$10 Gyr, Di Folco et al. 2007)
which represent $\sim$2$\%$ of all the Solar-type stars surveyed.
\emph{If} this group of objects corresponds to those that are currently
experiencing events similar to the LHB and the Hyades stars resemble
the Sun at 625\,Myr of age, then  the fact that none of the 67 Solar-type
stars in our Hyades sample has  a ``hot transient disk'' implies one of
two possibilities: (1) the likelihood of an  event similar to the LHB
is not significantly higher at $\sim$625\,Myr than it is at any other age,
or (2) events like the LHB are very short spikes with a duration much closer
to the lower limit of 10\,Myr suggested  by \citet{GoLeTs05} than to their
150\,Myr upper limit.  If the likelihood of a LHB-type event is approximately
constant with time, then a 2$\%$ incidence in the Solar neirghborwood
(median age $\sim$4000\,Myr) would imply a total duration of $\sim$80\,Myr.
However, if such an event is more likely to occur around an age of
$\sim$625\,Myr, then our non detections would only be consistent with a
much shorter duration. Thus, the implication of our results
on the LHB could depend on the age distribution of these ``hot transient
disks'', which still remains largely unconstrained since only 5 of such
examples are currently known.  Fortunately, as more \emph{Spitzer}
observations are reported, this distribution will become better constrained.

Understanding the debris disk phenomenon has been a high priority of the
\emph{Spitzer}'s mission. As a result, the number of debris disk studies
has increased dramatically over the last few years. Each of these studies
is providing new clues and constraints, from which it will eventually
emerge a much clearer picture of the evolution of debris disks
and its connection to the history of the Solar System.

\acknowledgements
We thank the anonymous referee for his/her many comments
and suggestions, which have helped to improve the paper significantly.
We are deeply grateful to Diane Paulson for helping to stimulate this work
and for valuable assistance in preparing the target lists.
This work is based on observations and on archival data obtained with the
Spitzer Space Telescope, which is operated by the Jet Propulsion Laboratory,
California Institute of Technology under a contract with NASA.
Support for this work was provided by NASA through an award issued by
JPL/Caltech and  through the Spitzer Space Telescope 
Fellowship Program. This publication makes use of the data products from the 
Two Micron All Sky Survey, which is a joint project of the University of Massachusetts 
and the Infrared Processing and Analysis Center/California Institute of
Technology, funded by the National Aeronautics and Space Administration and
the National Science Foundation.  

\clearpage
\bibliography{/home/syrah/rv/wdc/tex/bib/master.bib}

\newcommand{\bibfont}{\fontsize{10}{12}\selectfont} \newcommand{\noopsort}[1]{}
  \newcommand{\printfirst}[2]{#1} \newcommand{\singleletter}[1]{#1}
  \newcommand{\switchargs}[2]{#2#1}
\begin{thebibliography}{31}
\expandafter\ifx\csname natexlab\endcsname\relax\def\natexlab#1{#1}\fi

\bibitem[Augereau et al.(1999)]{1999A&A...348..557A} Augereau, J.~C., 
Lagrange, A.~M., Mouillet, D., Papaloizou, J.~C.~B., \& Grorod, P.~A.\ 
1999, \aap, 348, 557 

\bibitem[{Aumann {et~al.}(1984)Aumann, Beichman, Gillett, {de Jong}, Houck,
  Low, Neugebauer, Walker, \& Wesselius}]{AuBeGi84}
Aumann, H.~H., Beichman, C.~A., Gillett, F.~C., {de Jong}, T., Houck, J.~R.,
  Low, F.~J., Neugebauer, G., Walker, R.~G., \& Wesselius, P.~R. 1984, ApJ,
  278, L23

\bibitem[{Beichman {et~al.}(2005)Beichman, Bryden, Rieke, Stansberry, Trilling,
  Stapelfeldt, Werner, Engelbracht, Blaylock, Gordon, Chen, Su, \&
  Hines}]{BeBrRi05}

Beichman, C.~A., Bryden, G., Rieke, G.~H., Stansberry, J.~A., Trilling, D.~E.,
  Stapelfeldt, K.~R., Werner, M.~W., Engelbracht, C.~W., Blaylock, M., Gordon,
  K.~D., Chen, C.~H., Su, K.~Y.~L., \& Hines, D.~C. 2005, ApJ, 622, 1160

\bibitem[{{Beichman} {et~al.}(2006){Beichman}, {Bryden}, {Stapelfeldt},
  {Gautier}, {Grogan}, {Shao}, {Velusamy}, {Lawler}, {Blaylock}, {Rieke},
  {Lunine}, {Fischer}, {Marcy}, {Greaves}, {Wyatt}, {Holland}, \&
  {Dent}}]{BeBrSt06}

{Beichman}, C.~A., {Bryden}, G., {Stapelfeldt}, K.~R., {Gautier}, T.~N.,
  {Grogan}, K., {Shao}, M., {Velusamy}, T., {Lawler}, S.~M., {Blaylock}, M.,
  {Rieke}, G.~H., {Lunine}, J.~I., {Fischer}, D.~A., {Marcy}, G.~W., {Greaves},
  J.~S., {Wyatt}, M.~C., {Holland}, W.~S., \& {Dent}, W.~R.~F. 2006, ApJ, 652,
  1674

\bibitem[{Beichman {et~al.}(2007)Beichman, Fridlund, Traub, Stapelfeldt,
  Quirrenbach, \& Seager}]{BeFrTr07}

Beichman, C.~A., Fridlund, M., Traub, W.~A., Stapelfeldt, K.~R., Quirrenbach,
  A., \& Seager, S. 2007, in Protostars and Planets V, ed. B.~{Reipurth},
  D.~{Jewitt}, \& K.~{Keil}, 915--928

\bibitem[{Bryden {et~al.}(2006)Bryden, Beichman, Trilling, Rieke, Holmes,
  Lawler, Stapelfeldt, Werner, Gautier, Blaylock, Gordon, Stansberry, \&
  Su}]{BrBeTr06}

Bryden, G., Beichman, C.~A., Trilling, D.~E., Rieke, G.~H., Holmes, E.~K.,
  Lawler, S.~M., Stapelfeldt, K.~R., Werner, M.~W., Gautier, T.~N., Blaylock,
  M., Gordon, K.~D., Stansberry, J.~A., \& Su, K.~Y.~L. 2006, ApJ, 636, 1098

\bibitem[Cieza et al.(2007)]{2007ApJ...667..308C} Cieza, L., et al.\ 2007, 
\apj, 667, 308 

\bibitem[{{Chambers} \& {Lissauer}(2002)}]{ChLi02}
{Chambers}, J.~E., \& {Lissauer}, J.~J. 2002, in LPI Conference Abstracts, 1093

\bibitem[{Cochran {et~al.}(2002)Cochran, Hatzes, \& Paulson}]{CoHaPa02}
Cochran, W.~D., Hatzes, A.~P., \& Paulson, D.~B. 2002, AJ, 124, 565

\bibitem[{{Cohen}(2002)}]{Cohen02}
{Cohen}, B.~A. 2002, in LPI Conference Abstracts, 1984

\bibitem[Di Folco et al.(2007)]{2007A&A...475..243D} Di Folco, E., et al.\ 
2007, \aap, 475, 243 



\bibitem[Evans et al.(2003)]{2003PASP..115..965E} Evans, N.~J., et al.\ 
2003, \pasp, 115, 965


\bibitem[{{Evans} {et~al.}(2006){Evans}, {Harvey}, {Dunham}, {Mundy}, {Lai},
  {Chapman}, {Huard}, {Brooke}, \& {Koerner}}]{evans06}
{Evans}, N.~J., {Harvey}, P.~M., {Dunham}, M.~M., {Mundy}, L.~G., {Lai}, S.,
  {Chapman}, N., {Huard}, T., {Brooke}, T.~Y., \& {Koerner}, D.~W. 2006,   
  {Delivery of Data from the c2d Legacy Project: IRAC and MIPS (Pasadena,SSC)},
  Pasadena, SSC, http://ssc.spitzer.caltech.edu/legacy/original.html

\bibitem[Gautier et al.(2007)]{2007ApJ...667..527G} Gautier, T.~N., III, et 
al.\ 2007, \apj, 667, 527 

\bibitem[{Gomes {et~al.}(2005)Gomes, Levison, Tsiganis, \&
  Morbidelli}]{GoLeTs05}
Gomes, R., Levison, H.~F., Tsiganis, K., \& Morbidelli, A. 2005, Nature, 435,
  466

\bibitem[{Habing {et~al.}(2001)Habing, Dominik, {Jourdain de Muizon}, Laureijs,
  Kessler, Leech, Metcalfe, Salama, Siebenmorgen, Trams, \& Bouchet}]{HaDoJM01}
Habing, H.~J., Dominik, C., {Jourdain de Muizon}, M., Laureijs, R.~J., Kessler,
  M.~F., Leech, K., Metcalfe, L., Salama, A., Siebenmorgen, R., Trams, N., \&
  Bouchet, P. 2001, A\&A, 365, 545

\bibitem[{Hauschildt {et~al.}(1999)Hauschildt, Allard, \& Baron}]{HaAlBa99}
Hauschildt, P.~H., Allard, F., \& Baron, E. 1999, ApJ, 512, 377

\bibitem[{{Levison} {et~al.}(2001){Levison}, {Dones}, {Chapman}, {Stern},
  {Duncan}, \& {Zahnle}}]{LeDoCh01}
{Levison}, H.~F., {Dones}, L., {Chapman}, C.~R., {Stern}, S.~A., {Duncan},
  M.~J., \& {Zahnle}, K. 2001, Icarus, 151, 286

\bibitem[{Meyer {et~al.}(2006)Meyer, Hillenbrand, Backman, Beckwith, Bouwman,
  Brooke, Carpenter, Cohen, S.~Cortes, Crockett, Gorti, Henning, Hines,
  Hollenbach, Kim, Lunine, R.~Malhotra, Metchev, {Moro-Martin}, Morris, Najita,
  Padgett, Pascucci, Rodmann, Schlingman, Silverstone, Soderblom, Stauffer,
  Stobie, Strom, Watson, Weidenschilling, \& S.~Wolf}]{MeHiBa06}
Meyer, M.~R., Hillenbrand, L.~A., Backman, D., Beckwith, S., Bouwman, J.,
  Brooke, T., Carpenter, J., Cohen, M., S.~Cortes, S., Crockett, N., Gorti, U.,
  Henning, T., Hines, D., Hollenbach, D., Kim, J.~S., Lunine, J., R.~Malhotra,
  E.~M., Metchev, S., {Moro-Martin}, A., Morris, P., Najita, J., Padgett, D.,
  Pascucci, I., Rodmann, J., Schlingman, W., Silverstone, M., Soderblom, D.,
  Stauffer, J., Stobie, E., Strom, S., Watson, D., Weidenschilling, S., \&
  S.~Wolf, E.~Y. 2006, PASP, 118, 1690

\bibitem[{{Morbidelli} {et~al.}(2001){Morbidelli}, {Petit}, {Gladman}, \&
  {Chambers}}]{MoPeGl01}
{Morbidelli}, A., {Petit}, J.-M., {Gladman}, B., \& {Chambers}, J. 2001,
  Meteoritics and Planetary Science, 36, 371

\bibitem[{{Neukum} \& {Ivanov}(1994)}]{NeIv94}
{Neukum}, G., \& {Ivanov}, B.~A. 1994, in Hazards Due to Comets and Asteroids,
  359

\bibitem[{Paulson {et~al.}(2004)Paulson, Cochran, \& Hatzes}]{PaCoHa04}
Paulson, D.~B., Cochran, W.~D., \& Hatzes, A.~P. 2004, AJ, 127, 3579

\bibitem[{Paulson {et~al.}(2003)Paulson, Sneden, \& Cochran}]{PaSnCo03}
Paulson, D.~B., Sneden, C., \& Cochran, W.~D. 2003, AJ, 125, 3185

\bibitem[{Rieke {et~al.}(2005)Rieke, Su, Stansberry, Trilling, Bryden,
  Muzerolle, White, Gorlova, Young, Beichman, Stapelfeldt, \& Hines}]{RiSuSt05}
Rieke, G., Su, K.~Y.~L., Stansberry, J.~A., Trilling, D., Bryden, G.,
  Muzerolle, J., White, B., Gorlova, N., Young, E.~T., Beichman, C.~A.,
  Stapelfeldt, K.~R., \& Hines, D.~C. 2005, ApJ, 620, 1010

\bibitem[{{Ryder}(1990)}]{Ry90}
{Ryder}, G. 1990, EOS, 71, 313

\bibitem[{Schechter {et~al.}(1993)Schechter, Mateo, \& Saha}]{ScMaSa93}
Schechter, P.~L., Mateo, M., \& Saha, A. 1993, PASP, 105, 1342

\bibitem[{Song {et~al.}(2005)Song, Zuckerman, Weinberger, \&
  Becklin}]{SoZuWe05}
Song, I., Zuckerman, B., Weinberger, A.~J., \& Becklin, E.~E. 2005, Nature,
  436, 363

\bibitem[{Spangler {et~al.}(2001)Spangler, Sargent, Silverstone, Becklin, \&
  Zuckerman}]{SpSaSi01}
Spangler, C., Sargent, A.~I., Silverstone, M.~D., Becklin, E.~E., \& Zuckerman,
  B. 2001, ApJ, 555, 932

\bibitem[{Su {et~al.}(2006)Su, Rieke, Stansberry, Bryden, Stapelfeldt,
  Trilling, Muzerolle, Beichman, {Moro-Martin}, Hines, \& Werner}]{SuRiSt06}
Su, K.~Y.~L., Rieke, G.~H., Stansberry, J.~A., Bryden, G., Stapelfeldt, K.~R.,
  Trilling, D.~E., Muzerolle, J., Beichman, C.~A., {Moro-Martin}, A., Hines,
  D.~C., \& Werner, M.~W. 2006, ApJ, 653, 675

\bibitem[{{Tera} {et~al.}(1973){Tera}, {Papanastassiou}, \&
  {Wasserburg}}]{TePaWa73}
{Tera}, F., {Papanastassiou}, D.~A., \& {Wasserburg}, G.~J. 1973, in LPI Conf.
  Abstracts, 723

\bibitem[Trilling et al.(2007)]{2007arXiv0710.5498T} Trilling, D.~E., et 
al.\ 2007, ArXiv e-prints, 710, arXiv:0710.5498 

\bibitem[Weingartner \& Draine(2001)]{2001ApJ...548..296W} Weingartner, 
J.~C., \& Draine, B.~T.\ 2001, \apj, 548, 296 

\bibitem[{Wetherill(1975)}]{We75}
Wetherill, G.~W. 1975, in Lunar and Planetary Science Conference, 1539

\bibitem[{Wetherill(1977)}]{We77}
Wetherill, G.~W. 1977, in Lunar and Planetary Science Conference, 1

\bibitem[{Wyatt {et~al.}(2005)Wyatt, Greaves, Dent, \& Coulson}]{WyGrDe05}
Wyatt, M.~C., Greaves, J.~S., Dent, W.~R.~F., \& Coulson, I.~M. 2005, ApJ, 620,
  492

\bibitem[{Wyatt {et~al.}(2007{\natexlab{a}})Wyatt, Smith, Greaves, Beichman,
  Bryden, \& Lisse}]{WySmGr07}
Wyatt, M.~C., Smith, R., Greaves, J.~S., Beichman, C.~A., Bryden, G., \& Lisse,
  C.~M. 2007{\natexlab{a}}, ApJ, 658, 569

\bibitem[{Wyatt {et~al.}(2007{\natexlab{b}})Wyatt, Smith, Su, Rieke, Greaves,
  Beichman, \& Bryden}]{WySmSu07}
Wyatt, M.~C., Smith, R., Su, K.~Y.~L., Rieke, G.~H., Greaves, J.~S., Beichman,
  C.~A., \& Bryden, G. 2007{\natexlab{b}}, ApJ, 663, 365

\bibitem[{Yamashita {et~al.}(1993)Yamashita, Handa, Omodaka, Kitamura, Kawazoe,
  Hayashi, \& Kaifu}]{YaHaOm93}
Yamashita, T., Handa, T., Omodaka, T., Kitamura, Y., Kawazoe, E., Hayashi,
  S.~S., \& Kaifu, N. 1993, ApJ, 402, L65

\end{thebibliography}

\clearpage
\begin{deluxetable}{lcccrccccc}
\tabletypesize{\footnotesize}
\tablecaption{Sample of Hyades Stars \label{tab:sample}}
\tablehead{\colhead{Star Name}&\colhead{Ra}&\colhead{Dec}&\colhead{MIPS}&\colhead{IRAC}&\colhead{PID}&\colhead{Spectral}&\colhead{J}&\colhead{H}&\colhead{K$_S$}\\
\colhead{}&\colhead{(J2000.0)}&\colhead{(J2000.0)} &\colhead{AOR} &\colhead{AOR}  &\colhead{}   &\colhead{Type}    
&\colhead{(mag)} &\colhead{(mag)}&\colhead{(mag)}
}
\tablewidth{0pt}
\startdata
 BD+17 455  &  43.81800  &  17.89170      &   10853888  &   10865152  & 3371  & G7  & 7.56  & 7.21  & 7.18 \\
 BD+29 503  &  44.44480  &  29.66140      &   10841856  &   10854400  & 3371  & K0  & 7.38  & 7.00  & 6.91 \\
 HD 18632   &  45.01220  &   \phn7.74980  &   10842112  &   10854656  & 3371  & K2  & 6.32  & 5.95  & 5.84 \\
 HD 20430   &  49.35990  &   \phn7.65580  &   05403904  &   \nodata   &  148  & F8  & 6.29  & 6.05  & 5.99 \\
 BD+07 499  &  50.12200  &   \phn8.45440  &   12289280  &   12288512  & 3371  & K5  & 7.54  & 6.96  & 6.88 \\
 BD+23 465  &  53.20890  &  23.69240      &   10842624  &   10855168  & 3371  & K1  & 7.37  & 7.02  & 6.91 \\
 HIP 17766  &  57.04970  &   \phn7.14620  &   10843648  &   10856192  & 3371  & K6  & 8.27  & 7.62  & 7.51 \\
 BD+23 571  &  57.76320  &  23.90360      &   10843904  &   \nodata   & 3371  & K7  & 8.11  & 7.54  & 7.39 \\
 HD 286363  &  58.75610  &  12.48560      &   10854144  &   10856704  & 3371  & K4  & 8.18  & 7.69  & 7.57 \\
 HD 285252  &  58.77730  &  16.99840      &   10844160  &   10856960  & 3371  & K2  & 7.41  & 7.04  & 6.91 \\
\enddata
\tablecomments{The complete version of this table is in the electronic edition of
the Journal.  The printed edition contains only a sample.}
\end{deluxetable}


\begin{deluxetable}{lrrrrrr}
\tabletypesize{\footnotesize}
\tablecaption{IRAC Photometry \label{tab:IRAC}}
\tablehead{\colhead{Star name}&\colhead{Ra}&\colhead{Dec}&\colhead{3.6 $\mu$m Flux}&\colhead{4.5 $\mu$m Flux}&\colhead{5.8 $\mu$m Flux}&\colhead{8.0 $\mu$m Flux}\\
\colhead{}&\colhead{(J2000.0)}&\colhead{(J2000.0)}&\colhead{(mJy)}&\colhead{(mJy)}&\colhead{(mJy)}&\colhead{(mJy)}}
\tablewidth{0pt}
\startdata
BD+17 455 & 43.81795 &  17.89166   &  384 & 251 & 149 &  65 \\
BD+29 503 & 44.44457 &  29.66151   &  482 & 286 & 166 &  98 \\
HD 18632  & 45.01211 &   7.74987   & 1390 & 800 & 498 & 300 \\
BD+07 499 & 50.12181 &   8.45457   &  462 & 279 & 170 &  99 \\
BD+23 465 & 53.20876 &  23.69213   &  500 & 278 & 157 &  91 \\
HIP 17766 & 57.04950 &   7.14626   &  302 & 156 & 117 &  59 \\
HD 286363 & 58.75592 &  12.48568   &  250 & 143 &  95 &  54 \\
HD 285252 & 58.77703 &  16.99855   &  454 & 297 & 163 &  93 \\
BD+19 650 & 60.91259 &  19.45522   &  233 & 135 &  92 &  50 \\
HIP 19082 & 61.35694 &  19.44221   &  160 &  82 &  64 &  33 \\
\enddata
\tablecomments{[1]The complete version of this table is in the electronic edition of the Journal. The printed 
edition contains only a sample.]}
\tablecomments{[2] A 5$\%$ error should be adopted for all IRAC fluxes. 
As discussed in Section ~\ref{res_24}, even though the formal erros in our IRAC measurements 
are typically $<$1$\%$, there is a random  error floor to the best uncertainty possible
with our IRAC observing techniques and data reduction process of $\sim$5$\%$]}
\end{deluxetable}


\begin{deluxetable}{lrrrrrrrrrr}
\tablewidth{0pt}
\tablecaption{MIPS photometry and photospheric predictions\label{tab:MIPS}}
\tablehead{\colhead{Star name}&\colhead{Flux24}&\colhead{$\sigma$24}&\colhead{P24}&\colhead{Flux70}&\colhead{$\sigma$70}&\colhead{P70}&\colhead{Flux160}&\colhead{$\sigma$160}&\colhead{P160} \\
\colhead{}&\colhead{(mJy)}&\colhead{(mJy)}&\colhead{(mJy)}&\colhead{(mJy)}&\colhead{(mJy)}&\colhead{(mJy)}&\colhead{(mJy)}&\colhead{(mJy)}&\colhead{(mJy)}
}
\tablewidth{0pt}
\startdata
BD+17 455 &   9.72 &  0.095  &  10.4  & -0.97 & 3.00  & 1.22 &  21.1 & 12.50 & 0.23  \\
BD+29 503 &  12.2  &  0.093  &  14.0  & -5.07 & 2.82  & 1.64 &   7.8 & 30.60 & 0.31  \\
HD 18632  &  33.8  &  0.171  &  33.1  &  5.96 & 4.31  & 3.90 &  82.9 & 22.50 & 0.74  \\
HD 20430  &  28.1  &  0.097  &  29.3  & 11.10 & 4.58  & 3.45 &\nodata&\nodata&\nodata\\ 
BD+07 499 &  13.3  &  0.096  &  13.8  &  0.81 & 2.98  & 1.62 & -39.3 & 33.00 & 0.31  \\ 
BD+23 465 &  12.4  &  0.093  &  13.2  & -1.97 & 2.37  & 1.55 & -20.6 & 36.90 & 0.29  \\
HIP 17766 &   7.89 &  0.080  &   7.31 & -8.01 & 2.55  & 0.85 & -11.0 & 12.60 & 0.16  \\  
BD+23 571 &   8.23 &  0.086  &   9.15 &  3.07 & 2.55  & 1.08 & -15.7 & 17.70 & 0.20  \\
HD 286363 &   7.06 &  0.060  &   7.06 &  0.76 & 1.69  & 0.83 & -23.6 &  9.89 & 0.15  \\
HD 285252 &  12.2  &  0.092  &  12.4  & -2.04 & 3.22  & 1.46 & -53.5 & 25.60 & 0.27  \\
\enddata
\tablecomments{The complete version of this table is in the electronic edition of
the Journal.  The printed edition contains only a sample.}
\end{deluxetable}


\begin{deluxetable}{lccc}
\tablecaption{Disk Frequency Limits \label{tab:disklimits}}
\tablehead{\colhead{L$_{DUST}$/L$_*$ $\geq$}&\colhead{Number of stars}&\colhead{1-$\sigma$}&\colhead{2-$\sigma$}\\
\colhead{}&\colhead{above limit}&\colhead{}&\colhead{}}
\tablewidth{0pt}
\startdata                     
1$\times$10$^{-4}$ &  22  &  5.0$\%$ & 12.7 $\%$  \\
2$\times$10$^{-4}$ &  47  &  2.4$\%$ & 6.2$\%$  \\
5$\times$10$^{-4}$ &  64  &  1.8$\%$ & 4.6$\%$  \\
1$\times$10$^{-3}$ &  67  &  1.7$\%$ & 4.4$\%$  \\
\enddata
\end{deluxetable}


%
\begin{table*}
\label{dds_table}
\caption{Optically thin disk properties, for two different regimes of  minimal grain sizes $a_{\rm min}$.}
\label{nominal}
\begin{tabular}{lccccc}
\noalign{\smallskip}
\noalign{\smallskip}
\noalign{\smallskip}
\noalign{\smallskip}
\hline
\hline
\noalign{\smallskip}
& \multicolumn{2}{c}{$a_{\rm min} > 10\,\mu$m} & & \multicolumn{2} {c}{$a_{\rm min} < 0.5\,\mu$m}\\
\noalign{\smallskip}
\cline{2-3} \cline{5-6}
\noalign{\smallskip}
%
%
%
Star Name & $r_0$\,(AU) & $M_{\rm dust}$\,($10^{-3}$\,M$_{\oplus}$) &  & $r_0$\,(AU) & $M_{\rm dust}$\,($10^{-3}$\,M$_{\oplus}$)\\
\noalign{\smallskip}
\hline
\noalign{\smallskip}
HD 28226 &
$     46_{-       13}^{+     26} $ &
$     9.4_{-     6.9}^{+     24} $ &
 & 
$     475_{-    192}^{+ 24    } $ &
$      52_{-     19}^{+ 20    } $ \\
HD 28355 &
$      39_{-     9.2}^{+     11} $ & 
$      10_{-     7.5}^{+     16} $ &
 & 
$     383_{-     150}^{+ 116}$  &
$      57_{-      23}^{+  21}$   \\

\hline
\end{tabular}
%
%
\end{table*}

\clearpage

\clearpage

\clearpage

\clearpage

\clearpage

\clearpage

\clearpage

\clearpage

\clearpage

\end{document}